\documentclass[aps,prd,superscriptaddress,nofootinbib,amsmath,amsfonts,preprintnumbers,groupedaddress,showpacs,10pt,english]{revtex4-1}
\usepackage{amsmath}
\usepackage{amssymb}
\usepackage{babel}
\usepackage{wrapfig}
\usepackage{cancel}

\usepackage{relsize,exscale}
\makeatletter

\usepackage{array,multirow,graphicx}
\usepackage{dcolumn}
\usepackage{newlfont}
\usepackage{bm}
\usepackage[colorlinks,citecolor=blue,urlcolor=blue,linkcolor=blue]{hyperref}
\usepackage[figtopcap]{subfigure}
\usepackage{color}
\usepackage{verbatim}

\def\be{\begin{align}}
\def\ee{\end{align}}
\def\bea{\begin{eqnarray}}
\def\eea{\end{eqnarray}}
\def\bal{\begin{align}}
\def\eal{\end{align}}

\newcommand{\m}{\mathring}

\usepackage{scalerel}
\usepackage{tikz}
\usetikzlibrary{svg.path}
\definecolor{orcidlogocol}{HTML}{A6CE39}
\tikzset{
 orcidlogo/.pic={
 \fill[orcidlogocol] svg{M256,128c0,70.7-57.3,128-128,128C57.3,256,0,198.7,0,128C0,57.3,57.3,0,128,0C198.7,0,256,57.3,256,128z};
 \fill[white] svg{M86.3,186.2H70.9V79.1h15.4v48.4V186.2z}
 svg{M108.9,79.1h41.6c39.6,0,57,28.3,57,53.6c0,27.5-21.5,53.6-56.8,53.6h-41.8V79.1z M124.3,172.4h24.5c34.9,0,42.9-26.5,42.9-39.7c0-21.5-13.7-39.7-43.7-39.7h-23.7V172.4z}
 svg{M88.7,56.8c0,5.5-4.5,10.1-10.1,10.1c-5.6,0-10.1-4.6-10.1-10.1c0-5.6,4.5-10.1,10.1-10.1C84.2,46.7,88.7,51.3,88.7,56.8z};}}
\newcommand\orcid[1]{\href{https://orcid.org/#1}{\mbox{\scalerel*{
\begin{tikzpicture}[yscale=-1,transform shape]
\pic{orcidlogo};
\end{tikzpicture}
}{|}}}}
\graphicspath{{./}{Figs/}}
\begin{document}
\date{\today}

\title{ $f(Q)$ gravitational theory and its structure via red-shift}
\author{G.G.L. Nashed\orcid{0000-0001-5544-1119}}
\email{nashed@bue.edu.eg}
\affiliation {Centre for Theoretical Physics, The British University, P.O. Box
43, El Sherouk City, Cairo 11837, Egypt\\
Center for Space Research, North-West University, Potchefstroom 2520, South Africa.}

\begin{abstract}
We present a novel approach for reconstructing the
$f(Q)$ gravitational theory using parameterizations of the deceleration parameter or alternative options. This method enables the development of modified gravity scenarios that align with cosmological observations. We analyze two deceleration parameter models and one effective equation of state model from the literature. By leveraging the matter density parameter's asymptotic behaviour, we further refine the viable parameter space for these models. One tested model demonstrates that even slight modifications can yield viable cosmic scenarios, where the dark energy (DE) sector becomes dynamic and is entirely described instead of using a cosmological constant, we use modified gravity. These scenarios are qualitatively distinct yet quantitatively consistent with $\Lambda$CDM.
\end{abstract}
\maketitle
\section{Introduction}
Penrose and Hawking introduced the energy condition  in  general relativity to offer an explanation for the gravitational collapse-induced singularity, while adhering to non-negativity of local energy and considering the causal structure of the Universe \cite{Penrose:1964wq, Hawking:1966sx,Hawking:1966jv, Hawking:1967ju,1969CoASP...1....1H,Hawking:1970zqf}. Conventional matter does not match the observed evidence for the Universe's accelerated expansion \cite{SupernovaSearchTeam:1998fmf,SupernovaCosmologyProject:1998vns}, prompting the need for alternative energy known as dark energy. These model-independent evaluations all concur that dark energy, characterized by negative effective pressure, is the dominant component of the current cosmic matter. Regrettably, all efforts to detect dark energy in the universe through physical means have been unsuccessful. As a result, various curvature-based gravity theories were developed. $\mathrm{f(R)}$ gravity, $\mathrm{f(R,T)}$ gravity, $\mathrm{f(R,G)}$ gravity, and similar theories were proposed to address these issues through geometric interpretations rather than relying on the dark sector \cite{DeFelice:2010aj}. Albert Einstein himself introduced the metric teleparallel equivalent of GR (TEGR) and it is extensively researched, with the Levi Civita connection being substituted by a teleparallel connection based on torsion. A new theory in metric teleparallelism called $\mathrm{f(\mathbb{T})}$ theory was proposed to address the dark sector \cite{Ferraro:2006jd,Bahamonde:2021gfp,ElHanafy:2019zhr}.  Another type of teleparallel theory, known as symmetric teleparallelism, can be found in the existing literature \cite{Nester:1998mp}. The focus of this study is on the recently introduced $f\mathrm{(Q)}$ theories of gravity in symmetric teleparallelism, aiming to avoid dependencies on dark energy as proposed in \cite{BeltranJimenez:2017tkd}.

Einstein GR  is the conventional gravitational theory, which relies on curvature and the Einstein-Hilbert action \cite{CANTATA:2021ktz}. However, it is understood that gravity can also be represented in alternative ways, such as the torsional and non-metricity formulations, specifically with the teleparallel equivalent of GR (TEGR) \cite{Aldrovandi:2013wha}   and symmetric teleparallel equivalent of GR (STEGR) \cite{Nester:1998mp,BeltranJimenez:2017tkd}, respectively.  Combined, these three identical expressions make up the geometric trio of gravity \cite{BeltranJimenez:2019esp}. Changes in the curvature-based GR theory result in the popular $\mathrm{f(R)}$ gravity, $\mathrm{f(G)}$ gravity, Lovelock gravity, etc., as shown in \cite{Starobinsky:1980te, Capozziello:2002rd, Nojiri:2005jg, Lovelock:1971yv}. Additionally, $\mathrm{f(\mathbb{T})}$ gravity, an extension of TEGR, has been thoroughly examined and researched within the field of cosmology \cite{Cai:2015emx,Cai:2018rzd, Krssak:2018ywd,Nashed:2018qag,Yan:2019gbw, Huang:2022slc, Wang:2023qfm, Hu:2023juh}. Ultimately, adjustments involving the non-metricity scalar $Q$, such as expansions of the STEGR, result in $\mathrm{f(Q)}$ gravity. \cite{BeltranJimenez:2017tkd,Heisenberg:2023lru}. The cosmological implications of $\mathrm{f(Q)}$ gravity are quite fascinating, hence sparking a significant amount of recent research \cite{Khyllep:2021pcu,Mandal:2020buf,Barros:2020bgg,Lu:2019hra, De:2022wmj, Solanki:2022rwu, Lymperis:2022oyo,DAmbrosio:2021zpm,Li:2021mdp,Dimakis:2021gby, Hohmann:2021ast, Kar:2021juu,Wang:2021zaz,Quiros:2021eju,Mandal:2021bpd,Albuquerque:2022eac, Nashed:2022zyi,Capozziello:2022wgl, Capozziello:2022tvv,Dimakis:2022wkj,DAgostino:2022tdk, Narawade:2022cgb, Emtsova:2022uij,Bahamonde:2022cmz, Sokoliuk:2023ccw, De:2023xua, Dimakis:2023uib,Maurya:2023szc, Ferreira:2023awf,Capozziello:2023vne, Koussour:2023rly,Najera:2023wcw,Atayde:2023aoj, Paliathanasis:2023pqp,Bhar:2023zwi,Mussatayeva:2023aoa, Paliathanasis:2023kqs,Mandal:2023cag, Pradhan:2023oqo, Capozziello:2024vix,Bhar:2024vxk,Mhamdi:2024kgu}.

 In recent years, there have been several significant publications focusing on the $\mathrm{f(Q)}$ gravity theory and its implications for cosmology \cite{Mandal:2020buf,Lu:2019hra,Lin:2021uqa,De:2022wmj,Yang:2024tkw,Khyllep:2021pcu,De:2022shr,Solanki:2022rwu,Anagnostopoulos:2021ydo,Solanki:2022ccf,Beh:2021wva,
De:2022jvo,Barros:2020bgg,BeltranJimenez:2019tme,Frusciante:2021sio,Ferreira:2022jcd,Capozziello:2022wgl,Gadbail:2022jco,
Sarmah:2024fwi,Agrawal:2022vdg,Narawade:2022cgb,Narawade:2023nzv,Barros:2020bgg,Shabani:2023nvm,Nashed:2024ush,Maurya:2022wwa,Narawade:2023tnn,
Shabani:2023nvm,Narawade:2022cgb} and the mentioned references within. The corresponding energy condition were also discussed \cite{Mandal:2020lyq,Subramaniam:2023okn}. Nevertheless, besides Subramaniam et al. \cite{Subramaniam:2023okn}, all prior research was conducted exclusively within the spatially-flat FLRW model of the Universe, with the line element specifically using Cartesian coordinates. In this case, the $\mathrm{f(Q)}$ theory, known as the coincident gauge choice, was developed utilizing a vanishing affine connection.  The entire equation became simpler in this specific reference frame due to the fact that the covariant derivative changed into a partial derivative. Nevertheless, we had to make a trade-off by ensuring that the $\mathrm{f(Q)}$ theory and the $\mathrm{f(\mathbb{T})}$ theory were in alignment, resulting in matching Friedmann equations for pressure and energy density \cite{Jarv:2018bgs}. In  previous study \cite{Subramaniam:2023okn}, an analysis of this problem  is discussed by including a non-zero affine connection, but the analysis was still conducted within the context of a spatially flat FLRW spacetime.

The deceleration parameter $q(z)$ is a logical choice for describing the late rapid expansion, as any effective explanation of the universe's evolution must account for the transition from deceleration to acceleration to align with observations.  It is logical to use the deceleration parameter $q(z)$ to describe the recent accelerated expansion of the universe. A valid model of the universe's evolution must be able to transition from a phase of deceleration to one of acceleration, consistent with the observational evidence \cite{Cunha:2008ja,Cunha:2008mt,Nair:2011tg,Mamon:2016dlv,SupernovaSearchTeam:2004lze,Mamon:2017rri, Shafieloo:2007cs,Holsclaw:2011wi,Santos:2010gp,Holsclaw:2010nb,Crittenden:2011aa,Nair:2013sna,Sahni:2006pa}. Nevertheless, this method fails to offer a rationale for the dark energy's nature.  In the majority of cosmological studies, it is assumed by researchers that the observable Universe is spatially flat, meaning $k=0$.  Nevertheless, $k$ must be limited whenever the most recent observational data is accessible.  Hence, it would be ideal to consider the inclusion of the spatial curvature $k$. Some recent studies have thoroughly examined the impact of spatial curvature \cite{Chatzidakis:2022mpf,DiValentino:2019qzk,Yang:2022kho,Vagnozzi:2020rcz,Vagnozzi:2020dfn,Cruz:2018xzn,Dhawan:2021mel,Lai:2022sgp}. Studying the $\mathrm{f(Q)}$ theory in both open and closed type FLRW models with $k=\pm 1$ is undeniably beneficial. Until the study by Dimakis et al. \cite{Dimakis:2022rkd}, the primary difficulty was the intricate mathematical formalism of symmetric teleparallelism in a background spacetime, with little effort made to show a clear formulation in both open and closed type FLRW modes. The goal of this study is to place this kinematical method in the context of a revised theory of gravity that can also be tested further at the perturbation level. {A Big Bang Nucleosynthesis formalism and
observations were used in framework of $f({\mathrm Q})$ theory in order to extract constraints  on it \cite{Anagnostopoulos:2022gej}. They show \cite{Anagnostopoulos:2022gej} that $f({\mathrm Q})$ gravity can safely
pass the Big Bang Nucleosynthesis constraints which is not satisfied by some modified theories of gravity. The use of  Hubble data and Gaussian Processes  to reconstruct the dynamical connection function in $f({\mathrm Q})$ cosmology beyond the coincident gauge are investigated in \cite{Yang:2024tkw}. They showed that \cite{Yang:2024tkw}, in both cases and according to AIC and BIC information criteria the  inclusion of the non-coincident gauge is favored relative to $\Lambda$CDM paradigm.}

 The article is structured in the following way: Following the introduction, Section \ref{sec2} explains the fundamental mathematical framework of $\mathrm{f(Q)}$ theory, while Section \ref{sec3} discusses the derivation of $\mathrm{f(Q)}$ from a non-zero affine connection in a spatially curved FLRW Universe (positively and negatively curved).  Such connection coefficients involve a so-far unconstrained function of time, $\gamma(t)$. The Friedmann-like equations of energy and pressure for ordinary matter and for effective counterparts are also provided. EC expressions corresponding to $\mathrm{f(Q)}$ theory are presented in the brief Section \ref{sec5}. In the subsequent Section \ref{Sec3}, we will develop two formulation equations for $\mathrm{f(Q)}$ gravity using either the dark energy equation of state $\omega_{DE}(z)$ or the effective equation of state $\omega^{\text{Tot}}(z)$.  In Section \ref{Sec4}, we obtain the $\mathrm{f(Q)}$ gravity that gives rise to the $\Lambda$CDM model through the reconstruction presented in this study. Furthermore, we investigate two ways in which the deceleration parameter is described within the $\mathrm{f(Q)}$ theories. Moreover, we investigate a specific parametric expression of the effective equation of state \cite{Mukherjee:2016eqj}. Additionally, we analyze the outcomes of the three models. In conclusion, we provide a summary of the paper in Section \ref{sec6}.

\section{An overview of symmetrical teleparallelism}\label{sec2}
The theory of symmetric teleparallel gravity was constructed using an affine connection, ${\mathrm \Gamma^\alpha_{\,\,\, \beta\gamma}}$, which is considered general and defined as follows:
\begin{equation} \label{connc}
{\mathrm \Gamma^\lambda{}_{\mu\nu} = \mathring{\Gamma}^\lambda{}_{\mu\nu}+L^\lambda{}_{\mu\nu}}\,.
\end{equation}
In the symmetrical teleparallelism theory, gravity is governed by the non-metricity of the underlying geometry, which has no curvature and zero torsion.  The first step is to establish the definition of the non-metricity tensor that has the form:
\begin{equation} \label{Q tensor}
{\mathrm {Q_{\lambda\mu\nu} = \nabla_\lambda g_{\mu\nu}}} \,.
\end{equation}
There are two potential traces for the non-metricity tensor, are defined as follows:
\[
{\mathrm Q_{\lambda}=Q_{\lambda\mu\nu}g^{\mu\nu}, \qquad \qquad \tilde Q_{\nu}=Q_{\lambda\mu\nu}g^{\lambda\mu}}\,.
\]
The  tensors $L^\lambda{}_{\mu\nu}$ and  $P^\lambda{}_{\mu\nu}$ are respectively define the disformation and the superpotential and are defines as:
\begin{equation} \label{L}
{\mathrm L^\lambda{}_{\mu\nu} = \frac{1}{2} (Q^\lambda{}_{\mu\nu} - Q_\mu{}^\lambda{}_\nu - Q_\nu{}^\lambda{}_\mu)} \,,
\end{equation}
\begin{equation} \label{P}
{\mathrm P^\lambda{}_{\mu\nu} = \frac{1}{4} \left( -2 L^\lambda{}_{\mu\nu} + Q^\lambda g_{\mu\nu} - \tilde{Q}^\lambda g_{\mu\nu} -\frac{1}{2} \delta^\lambda_\mu Q_{\nu} - \frac{1}{2} \delta^\lambda_\nu Q_{\mu} \right)} \,.
\end{equation}
We are examining the scalar quantity of non-metricity, which has the form:
\begin{equation} \label{Q}
{\mathrm Q=Q_{\lambda\mu\nu}P^{\lambda\mu\nu}= \frac{1}{4}( 2Q_{\lambda\mu\nu}Q^{\mu\lambda\nu} -Q_{\lambda\mu\nu}Q^{\lambda\mu\nu} +Q_\lambda Q^\lambda -2Q_\lambda \tilde{Q}^\lambda)}\,.
\end{equation}

However, like GR, symmetric teleparallelism also encounters the challenging issue known as the 'dark' problem. Therefore, a modified $f(Q)$ gravity theory has been proposed, similar to the introduction of modified $f(R)$ theory, to expand upon GR. The action of this modified gravitational theory is given by:
\begin{equation}\label{action}
{\mathrm S = \frac1{2\kappa}\int f(Q) \sqrt{-g}\,d^4 x
+\int \mathcal{L}_M \sqrt{-g}\,d^4 x}\,,
\end{equation}
where $f(Q)$ is an arbitrary function of the non-metricity scalar, i.e., $Q$. Variation of  Eq.~\eqref{action}
w.r.t. the metric yields the   equation of motions as \cite{BeltranJimenez:2017tkd}:
\begin{equation} \label{FE}
{\mathrm f_Q \m{G}_{\mu\nu}+\frac{1}{2} g_{\mu\nu} (f_QQ-f) + 2f_{QQ} P^\lambda{}_{\mu\nu} \m{\nabla}_\lambda Q = \kappa T^{m}_{\mu\nu}}\,,
\end{equation}
where $\mathrm {\m{G}_{\mu\nu}}$ is the Einstein tensor. Now let us transforms Eq.~(\ref{FE}) into an equivalent form of GR as:
\begin{align}\label{FEeq}
    \m{G}_{\mu\nu}= \frac{\kappa T^{eff}_{\mu\nu}}{f_Q} =\frac {\kappa T^{m}_{\mu\nu}}{f_Q}+T^{DE}_{\mu\nu}\,,
\end{align}
where \begin{align}
{\mathrm T^{DE}_{\mu\nu}=\frac{1}{\kappa f_Q}\left[\frac{1}{2}g_{\mu\nu}(f-Qf_Q)-2f_{QQ}\mathring{\nabla}_\lambda QP^\lambda_{\mu\nu}\right]}\,, \end{align} refers to the extra terms resulting from the geometric alteration of the theory of gravity in this case. This can be easily imagined as the element that functions as a sort of imaginary dark force.

\section{Homogeneous and isotropic model in the frame of $f(Q)$}\label{sec3}
The metric for an FLRW spacetime with spatial curvature and isotropic homogeneity is defined as:
\begin{align}\label{le}
{\mathrm ds^2 = -\mathrm{d} t^2
+R\left(t\right)^{2}\left( \frac{dr^2}{1-kr^2} +r^2\mathrm{d}\theta^2+r^2\sin^2\theta\mathrm{d} \phi^2\right)\,, \quad \mbox{where} \quad k=\pm1\,,}
\end{align}
where ${\mathrm R(t)}$ is the scale factor.
In the frame of this spacetime, the compatible connection was deliberated in \cite{Dimakis:2022rkd} as:
\begin{align}\label{con}
{\mathrm \Gamma^t{}_{tt}}=&-\mathrm{\frac{k+\dot\gamma}{\gamma}} \,,
	\quad 					{\mathrm \Gamma^t{}_{rr}}=\mathrm {\frac{\gamma}{1-kr^2}\,},
	\quad 					{\mathrm \Gamma^t{}_{\theta\theta}=\gamma r^2}\,,
	\quad						{\mathrm \Gamma^t{}_{\phi\phi}=\gamma r^2\sin^2\theta}\,,	\quad {\mathrm \Gamma^r{}_{tr}}=-\mathrm {\frac{k}{\gamma}}\,,							\notag\\
{\mathrm \Gamma^r{}_{rr}}=&\mathrm{\frac{kr}{1-kr^2}}\,,
	\quad		{\mathrm \Gamma^r{}_{\theta\theta}=-(1-kr^2)r}\,,
	\quad		{\mathrm \Gamma^r{}_{\phi\phi}=-(1-kr^2)r\sin^2\theta}\,,	\quad {\mathrm \Gamma^\theta{}_{t\theta}=-\frac{k}{\gamma}}\,,											 \notag\\
	\quad		{\mathrm \Gamma^\theta{}_{r\theta}}=&\mathrm {\frac1r}\,,
	\quad		{\mathrm \Gamma^\theta{}_{\phi\phi}=-\cos\theta\sin\theta}\,,\quad {\mathrm \Gamma^\phi{}_{t\phi}}=-\mathrm {\frac k\gamma}\,,
	\quad 	{\mathrm \Gamma^\phi{}_{r\phi}=\frac1r}\,,
	\quad 	{\mathrm \Gamma^\phi{}_{\theta\phi}=\cot\theta}\,.
\end{align}

where $\gamma\equiv \gamma(t)$.
The non-metricity scalar $Q$ can be determined by using equation (\ref{Q}) to give:
\begin{equation}\label{Qf}
 {\mathrm   Q(t)=-3\left[2H^2+\left(\frac{3k}{\gamma}-\frac{\gamma}{R^2}\right)H-\frac{2k}{R^2}-k\frac{\dot{\gamma}}{\gamma^2}-\frac{\dot{\gamma}}{R^2}\right]}\,.
\end{equation}
Now let's examine regular matter that behaves like a perfect fluid, with its stress-energy tensor ${\mathrm T^{m}_{\alpha\beta}}$ given as:
\begin{align}
   {\mathrm T^{m}_{\alpha\beta}=(p^{m}+\rho^{m})u_\alpha u_\beta+p^{m}g_{\alpha\beta}}\,.
\end{align}
In this case, the fluid's four-velocity is represented by the unit vector $u^{\mu}$. The FRW  equations can be derived from the field equation (\ref{FE}) that gives:
\begin{align}\label{rho}
{\mathrm \rho^{m}=\frac12f+\left(3H^2+3\frac k{R^2}-\frac12Q\right)f_Q+\frac32\dot Q\left(-\frac k\gamma-\frac\gamma{R^2}\right)f_{QQ}}\,.
\end{align}
\begin{align}\label{p}
{\mathrm p^{m}=-\frac12f+\left(-3H^2-2\dot H-\frac k{R^2}+\frac12Q\right)f_Q
        +\dot Q\left(-2H-\frac32\frac k\gamma+\frac12\frac\gamma{R^2}\right)f_{QQ}}\,,
\end{align}
where the symbols ${\mathrm p^{m}}$ and ${\mathrm \rho^{m}}$ represent the pressure and energy density, respectively.
The total  energy density and pressure yield the form:
\begin{align}\label{rho_eff}
&{\mathrm \rho^{Tot}=\rho^m+\frac{1}{2}(Qf_Q-f)+\frac{3}{2}\dot{Q}f_{QQ}\left(
        \frac{\gamma}{R^2}+\frac{k}{\gamma}\right)}\,\\
&{\mathrm p^{Tot}=p^m-\frac{1}{2}(Qf_Q-f)
   -\frac12\dot{Q}f_{QQ}\left(\frac{\gamma}{R^2}-\frac{3k}{\gamma}-4H \right)}\,.
\label{p_eff}
\end{align}

where $\dot {\mathrm Q}\equiv \frac{d{\mathrm Q}}{dt}$, \quad ${\mathrm f_Q\equiv \frac{\partial f}{\partial Q}}$, \quad  ${\mathrm f_{QQ}\equiv \frac{\partial^2 f}{\partial Q^2}}$.

\section{Analysis focused on a particular model}\label{sec5}

In this section, we consider the extension of GR, in terms of  $f(Q)$. In spherical coordinates, with the line element given by (\ref{le}) and spatial curvature $k=0$, we set $\gamma(t)=0$ \cite{Dimakis:2022rkd}. he FRW equations can be expressed by using the relation (\ref{Qf}) and substituting Eq. (\ref{le}) into the field equations (\ref{FE}) in terms of $f(Q)$ as\footnote{Equations \eqref{FR2H} reduce to the symmetric teleparallel equivalent of GR when $f(Q)\equiv Q$.}:
\begin{eqnarray}
 {\mathrm \rho^{m}} &=& ~\mathrm {\frac{1}{2\kappa}\left(f-H f_{H}\right)}\,, \label{FR1H}\\
\nonumber {\mathrm p^{m}}    &=& -\mathrm{\frac{1}{2\kappa}\left(f-H f_{H}-\frac{1}{3}\dot{H} f_{HH}\right)}\equiv \mathrm{\frac{1}{6\kappa}\dot{H} f_{HH}-\rho^{m}}, \label{FR2H}
\end{eqnarray}
where ${\mathrm f_Q=\frac{df}{dQ}\equiv- \frac{f_H}{12H}}$ with $\mathrm {f_H=\frac{df}{dH}}$ similar for ${\mathrm f_{QQ}=\frac{d^2f}{dQ^2}}$ with ${\mathrm f_{HH}=\frac{d^2f}{dH^2}}$.

To close the system, it is necessary to select an equation of state (EoS) to link $\rho^{m}$ and $p^{m}$. In the simplest barotropic case, the system mentioned generates the applicable dynamic equation, in which $\mathrm{p^m \equiv p^m(\rho^m) = w_m \rho^m}$.
\begin{align}\label{phase-portrait}
    {\mathit {\dot{H}=3(1+w_{m})\left[\frac{f(H)-H f_{H}}{f_{HH}}\right]=\mathcal{F}(H)}}\,.
\end{align}
Because  $\dot{H}$ depends only on $H$, as evident from the previous relation, the phase portrait of any ${\mathrm f(Q)}$ theory in a flat FRW background is described by Eq. (\ref{phase-portrait}).  This study examines the later stages of cosmic evolution, assuming a universe predominantly made up of baryons with $w_{m}=0$ before transitionary to dark energy dominance.

Modified gravity theories should be regarded as extensions of general relativity due to the remarkable results they yield. Incorporating Einstein's gravity along with correction terms from higher-order $\mathrm{f(Q)}$ symmetric teleparallel gravity into the field equations proves beneficial. As a result, the modified FRW equations are expressed as follows:
\begin{eqnarray}
{\mathrm {H}^2= \frac{\kappa}{3} \left( \rho^{m}+  \rho^{ Q} \right)}&\equiv& \mathrm{\frac{\kappa}{3} \rho^{Tot}}, \label{MFR1}\\
{\mathrm 2 \dot{{H}} + 3{H}^2= - \kappa  \left( p^{m}+p^{ Q }\right)}&\equiv& \mathrm{-\kappa p^{Tot}}\,.\label{MFR2}
\end{eqnarray}
In this scenario, the  pressure and energy-density of the symmetric teleparallel  of ${\mathrm f(Q)}$ are determined by:
\begin{eqnarray}
{\mathrm \rho^{Q}(H)}&=&\mathrm{\frac{1}{2\kappa}\left(H f_{H}-f(H)+6H^{2}\right)},\label{rhoT}\\
{\mathrm p^{Q}(H)}&=&\mathrm{-\frac{1}{6\kappa}\dot{H}\left(12+f_{HH}\right)-\rho^{Q}(H)}.\label{pT}
\end{eqnarray}
{ When $\mathrm{f(Q)}=Q$ which is the GR limit, then ${\mathrm \rho^Q}$ and ${\mathrm p^Q}$ are both equal to zero.
In cases  ${\mathrm f(Q)}$ is not linear, the non-metricity scalar ${\mathrm Q}$ corresponding to ${\mathrm f(Q)}$ may act as the dark energy.}  In the barotropic situation, the non-metricity scalar will possess an equation of state  as
\begin{equation}\label{torsion_EoS}
 {\mathit  \omega_{DE}=\omega _{Q}(H)=-1-\frac{1}{3}\frac{\dot{H}(12+f_{HH})}{6H^2-f(H)+Hf_{H}}}\,.
\end{equation}
Let $\rho_{c}$ stand for the critical density and $i$ for the species component ($\equiv \rho^{\text{Tot}}$). The density parameters are defined as $w_{i} = \frac{\rho_{i}}{\rho_{c}}$. Consequently, the FRW equation (\ref{MFR1}) can be expressed in its dimensionless form as follows:
\begin{equation}\label{dimless_FR}
 \mathit{ w_m+w_Q=1}\,,
\end{equation}
with ${\mathrm w_m}$ being  the matter density parameter, calculated as $\mathrm {\frac{\kappa \rho^m}{3H^2}}$, while ${\mathrm w_Q}$ is the non-symmetric density parameter, calculated as $\frac{\kappa\rho_Q}{3H^2}$. In order to uphold the conservation law, the continuity equations are obtained by minimally coupling the matter field and the non-symmetric.
\begin{eqnarray}
  \mathrm {\dot{\rho}^{m}+3H \rho^{m}} &=& 0, \label{sc-continuity}\\
 \mathrm {\dot{\rho}^Q+3H(\rho^{Q}+p^{Q})} &=& 0. \label{tor-continuity}
\end{eqnarray}
Defining the total EoS parameter can also be beneficial and in this case it takes the form:
\begin{equation}\label{eff_EoS}
{\mathrm \omega^{Tot}\equiv \frac{p^{Tot}}{\rho_{Tot}}=-1-\frac{2}{3}\frac{\dot{H}}{H^2}}.
\end{equation}
Since they are interconnected, the total equation of state (EoS) parameter can be used as a substitute for the deceleration parameter ${\mathit q}$ as:
\begin{equation}\label{deceleration}
\mathrm  {q \equiv -1-\frac{\dot{H}}{H^{2}}=\frac{1}{2}\left(1+3\omega^{Tot}\right)}.
\end{equation}
Cosmological observations have revealed that the universe transitioned from a decelerating phase to an accelerating phase of expansion several billion years ago.  Since the discovery of the universe's transition from deceleration to acceleration, the deceleration parameter $q$ has been extensively utilized to chronicle the history of the universe up to the present day.  Thus, some people used different parametric versions of ${\mathrm q}$ \cite{Cunha:2008ja,Nair:2011tg,Mamon:2016dlv,Cunha:2008mt,Santos:2010gp,Mamon:2017rri,SupernovaSearchTeam:2004lze}, while others used non-parametric versions of ${\mathrm q}$ \cite{Holsclaw:2010nb,Crittenden:2011aa,Nair:2013sna,Holsclaw:2011wi,Shafieloo:2007cs,Sahni:2006pa}.  Nonetheless, a theory of gravity must be used to define these conditions. The subsequent section outlines a process for reconstructing ${\mathrm f(Q)}$ gravity models based on a specified expression for the deceleration parameter ${\mathrm q(z)}$, with $z$ representing the redshift. This approach can also incorporate additional parameters like the total equation of state ${\mathrm w^{Tot}(z)}$. By employing this method, we can delve deeper into the examination of the free parameters involved in these formulations. Additional understanding can be achieved by connecting these parameters to other cosmological measures, including the matter density parameter $\Omega_m$ and the dark energy equation of state. This reconstruction technique provides a powerful tool for understanding the dynamics of ${\mathrm f(Q)}$ gravity within the framework of observational cosmology.
\section{Method of reconstructing ${\mathrm f(Q)}$ in symmetric teleparallel theory}\label{Sec3}
Given the relationship $z = \frac{R_{0}}{R} - 1$, where $R_{0}$ is the scale factor at the present time and $R$ is the scale factor at any other time, it becomes straightforward to employ the redshift $z$ as the independent variable in cosmological analysis. This makes it convenient to express the evolution of cosmological parameters and equations in terms of $z$. Under these conditions, we proceed to formulate our equations with $z$ as the central variable, enabling a clear examination of how the universe's properties change over time and write \begin{equation}\label{redsh-Hubble}
{\mathit {\dot{H}=-(1+z)H H'}}\,,
\end{equation}

In terms of the redshift parameter $z$, the prime indicates differentiation with respect to $z$. Using Eqs.~(\ref{deceleration}) and (\ref{redsh-Hubble}), ${\mathrm H(z)}$ can be expressed as follows:
\begin{equation}\label{Hubble-deceleration}
 {\mathrm   H(z)=H_{0}\exp\left(\int_{0}^{z}\frac{1+q(\tilde{z})}{1+\tilde{z}}d\tilde{z}\right)}\,,
\end{equation}
where ${\mathrm H_{0}}$ is equal to ${\mathrm H}$ at   ${\mathrm z}=0$. Conversely, when     ${\mathrm z}$ is used as   independent variable, we derive the following:
\begin{equation}\label{z-trans}
 {\mathit    f(H(z))=f(z)}, \qquad  {\mathrm f_{H}=\frac{f'}{H'}}, \qquad {\mathrm f_{HH}=\frac{f'' H'- f' H''}{H'^{3}}}\,.
\end{equation}
By inserting Eqs.~(\ref{redsh-Hubble}) and (\ref{z-trans}) into the phase portrait ${\mathrm f(Q)}$ described in Eq.~(\ref{phase-portrait}), we can evaluate $H(z)$ as:
\begin{equation}\label{phase-portrait-z}
 {\mathit    H(z)=H_{0}\exp\left(\int_{0}^{z}\frac{f'(\tilde{z})}{f(\tilde{z})+f_{0}(1+\tilde{z})^{3}} d\tilde{z}\right)}\,,
\end{equation}
with ${\mathit f_{0}}$  being a constant that gets incorporated into the equation.  Through the utilization of Eqs.~(\ref{redsh-Hubble}) and (\ref{z-trans}) into Eq.~(\ref{FR1H}), we determine the   density of the matter as a function of the red-shift as:

\begin{equation}\label{matter-density}
    \mathrm{\rho^{m}(z)=\frac{1}{2\kappa}\left(f(z)-\frac{H}{H'}f'\right)=-\frac{f_{0}}{2\kappa}(1+z)^{3}}\,.
\end{equation}
Alternatively, the matter density can be determined by solving the matter continuity equation (\ref{sc-continuity}), where the present matter density is $\rho_{m,0}$ and $\rho^m = \frac{\rho_{m,0}}{R^{3}} = \rho_{m,0}(1+z)^3$. The difference from Eq.~(\ref{matter-density}) is given by $f_0 = -2\kappa \rho_{m,0}$. Currently ($\mathrm {z=0}$), we have $w_{m,0}=\frac{\rho_{m,0}}{\rho_{c,0}}=\frac{\kappa\rho_{m,0}}{3 H_0^2}$, which results in:
\begin{equation}\label{f0}
\mathrm {f_{0}=-6w_{m,0}H_{0}^{2}}\,.
\end{equation}
The Planck CMB results show that $w_{m,0}h^2=0.1426 \pm 0.0020$ when using the $\Lambda$CDM model with Planck TT+lowP likelihood, where $h=H_0/100$ km/s/Mpc \cite{Planck:2015fie}. This directly determines the value of the constant to be $f_{0}=-8556$.

Modified $\mathrm{f(Q)}$ gravity and the deceleration parameter $\mathrm{q(z)}$ have a similar connection with the Hubble function $\mathrm{H(z)}$, according to  Eqs.~(\ref{Hubble-deceleration}) and (\ref{phase-portrait-z}) which yields:
\begin{equation}\label{deceleration-f(z)}
{\mathit {q(z)=\frac{(1+z)f'}{f(z)-6w_{m,0}H_0^2(1+z)^{3}}}-1}\,.
\end{equation}
As discussed earlier, the nature of the cosmic expansion rate is directly linked to the deceleration parameter. Therefore, various parameterization forms of the decelerations have been proposed in research to depict the cosmic evolution \cite{SupernovaSearchTeam:2004lze,Cunha:2008ja,Cunha:2008mt,Santos:2010gp,Nair:2011tg,Mamon:2016dlv,Mamon:2017rri, Shafieloo:2007cs,Holsclaw:2010nb,Holsclaw:2011wi,Crittenden:2011aa,Nair:2013sna,Sahni:2006pa}. Therefore, it can be concluded that Eq.~(\ref{deceleration-f(z)}) serves as a resource for creating feasible cosmic scenarios in the framework of  ${\mathrm f(Q)}$ gravitational theory.


Instead, Eq.~(\ref{deceleration-f(z)}) can be used in the integration to derive  $\mathrm{f(z)}$ as:
\begin{equation}\label{Reconstruction1}
 {\mathit{   f(z)=-6w_{m,0}H_0^2e^{{\textstyle{\int_0^z}}\frac{1+q(\tilde{z})}{1+\tilde{z}}d\tilde{z}}{\displaystyle{\int_0^z}}
    \frac{(1+\tilde{z})^{2}(1+q(\tilde{z}))}{e^{{\textstyle{\int_0^z}}\frac{1+q(\tilde{z})}{1+\tilde{z}}d\tilde{z}}}d\tilde{z}}}\,.
\end{equation}
Therefore, with a specific parametrization of $\mathrm{q(z)}$, Eq. (\ref{Reconstruction1}) allows for the creation of the corresponding ${\mathrm f(Q)}$ theory, followed by the calculation and comparison of other key parameters with observational data to assess the credibility of $\mathrm{f(Q)}$ theory.
 In the present study,  Eq.~(\ref{Reconstruction1}) is employed  to analyze the $\mathrm {f(Q)}$ gravitational theory and its associated parametrization $\mathrm {q(z)}$. Furthermore, we can swap $\mathrm{q(z)}$ with $\mathrm{w^{Tot}(z)}$ using (\ref{deceleration}), allowing for the reconstruction of $\mathrm {f(Q)}$ gravity with various parameterizations of $\mathrm {w^{Tot}(z)}$ as well.Here, we encounter a different reconstruction equation, given by:
\begin{eqnarray}
 { \mathit{ f(z)}=\mathrm{-9w_{m,0}H_0^2e^{\frac{3}{2}{\textstyle{\int_0^z}\frac{1+w^{Tot}(\tilde{z})}{1+\tilde{z}}}d\tilde{z}}} \mathrm{\int_0^z
   \frac{(1+\tilde{z})^{2}(1+w^{Tot}(\tilde{z}))}
    {e^{\frac{3}{2}{\textstyle{\int_0^z \frac{1+w_{eff}(\tilde{z})}{1+\tilde{z}}}} d\tilde{z}}} d\tilde{z}}}\,.\label{Reconstruction2}
\end{eqnarray}
To encompass additional reconstruction techniques, we consider scenarios where parameterizations are provided for the EoS of the field of dark energy. Eqs.~(\ref{redsh-Hubble}) and (\ref{z-trans}) are substituted in Eq.~(\ref{torsion_EoS}) to obtain:
\begin{eqnarray}
\mathrm{\omega_{Q}(z)=-1+\frac{1}{3}(1+z)H\frac{12H'^{3}+f''H'-f'H''}{(6H^{2}-f)H'^{2}+HH'f'}}\,.\label{wQ(z)}
\end{eqnarray}
Equation (\ref{wQ(z)}) can be employed to recreate $\mathrm{f(z)}$ using a specified parametrization of the dark energy  EoS $\mathrm{w_Q(z)}$. By substituting Eq.~(\ref{phase-portrait-z}) in Eq.~(\ref{wQ(z)}), we can obtain a novel formulation given by:
\begin{widetext}
\begin{equation}\label{Reconstruction3}
{\mathit{ \omega_Q(z)=\frac{\left[f(z)-6w_{m,0}H_0^2(1+z)^3-\frac{2}{3}(1+z)f'(\tilde{z})\right]
    e^{2{\textstyle{\int_0^z}}\frac{f'(\tilde{z})}{f(\tilde{z})-6w_{m,0}H_0^2(1+\tilde{z})^3} d\tilde{z}}}
    {\left[f(z)-6w_{m,0}H_0^2(1+z)^3\right]\left[w_{m,0}(1+z)^3-e^{2{\textstyle{\int_0^z}}\frac{f'(\tilde{z})}{f(\tilde{z})
    -6w_{m,0}H_0^2(1+\tilde{z})^3}d\tilde{z}}\right]}}}\,.
\end{equation}
\end{widetext}
By inserting Eqs.~(\ref{Hubble-deceleration}) and (\ref{Reconstruction1}) in Eq.~(\ref{wQ(z)}), we uncover a significant relationship between the dark energy EoS and the deceleration parameter, expressed as follows:
\begin{equation}\label{wQ-deceleration}
  {\mathit{  \omega_{Q}(z)  =\frac{\left(1-2q(z)\right)e^{2{\textstyle{\int_0^z}} \frac{1+q(\tilde{z})}{1+\tilde{z}} d\tilde{z}}}{3\left(w_{m,0}(1+z)^3- e^{2{\textstyle{\int_0^z}} \frac{1+q(\tilde{z})}{1+\tilde{z}} d\tilde{z}}\right)}}}\,,
\end{equation}
Once more, we can use $q(z)$ and $\omega^{Tot}(z)$ mutually using Eq.~(\ref{deceleration}). This enables us to determine the following relationship between the entire equation of state parameters and the dark energy:
\begin{equation}\label{wT-weff}
{\mathit{    \omega_{Q}(z)  =-\frac{\omega^{Tot}(z)e^{3{\textstyle{\int_0^z} \frac{1+w_{eff}(\tilde{z})}{1+\tilde{z}} d\tilde{z}}}}{w_{m,0}(1+z)^3- e^{3{\textstyle{\int_0^z} \frac{1+\omega_{Tot}(\tilde{z})}{1+\tilde{z}} d\tilde{z}}}}}}\,,
\end{equation}

To conclude this section, we establish a connection between the deceleration parameter and another essential cosmological parameter. It makes it possible to use the matter density parameter $\mathrm{w_{m}(z)}$ to examine assumed forms of $q(z)$. (\ref{matter-density}) is used to write:
\begin{equation}\label{matter-density-parameter}
{\mathit {w_{m}(z)=\frac{fH'-Hf'}{6H' H^{2}}=w_{m,0}(1+z)^{3}\, e^{-2\textstyle{\int_0^z \frac{1+q(\tilde{z})}{1+\tilde{z}}d\tilde{z}}}}}\,.
\end{equation}
Furthermore, the dark symmetrical teleparallel counterpart is subsequently provided by:
\begin{equation}\label{torsion-density-parameter}
{\mathit{w_{Q}=1-w_{m}=1-w_{m,0}(1+z)^{3}\, e^{-2\textstyle{\int_0^z \frac{1+q(\tilde{z})}{1+\tilde{z}}d\tilde{z}}}}}\,.
\end{equation}

 In the next part, we will utilize these equations to investigate various parametric forms in $\mathrm{f(Q)}$ gravity.

\section{Applications}\label{Sec4}
Inspired by the finding results in Sec. \ref{Sec3}, to develop the corresponding $\mathrm{f(Q)}$ gravity theory and explore potential deviations from the $\Lambda$CDM model, we introduce three parameterizations: When using redshift ${\mathrm z}$ as the independent variable, we derive two expressions for the deceleration parameter $\mathrm{q(z)}$, along with one for the effective EoS parameter $\mathrm{\omega^{Tot}(z)}$:
\subsection{$\Lambda$CDM flat model}\label{Sec4.1}
 The Hubble parameter's evolution is shown in this framework as follows:
\begin{equation}\label{HLCDM}
 {\mathit{   H(z)=H_0\sqrt{w_{m,0}(1+z)^3+w_{\lambda,0}}}} \, ,
\end{equation}
$w_{\lambda,0}$ represents one minus $w_{m,0}$, which indicates the current value of the dark energy density parameter.  By plugging Eq.~(\ref{HLCDM}) in Eq.~(\ref{deceleration}) and considering Eq.~(\ref{redsh-Hubble}), we express the deceleration parameter for $\Lambda$CDM as:
\begin{equation}\label{qLCDM}
\mathrm{    q(z)=\frac{3}{2}\frac{w_{m,0}(1+z)^3}{w_{\lambda,0}+w_{m,0}(1+z)^3}}-1\,.
\end{equation}
At high redshifts, when $\mathrm{(1+z)^3}$ is much greater than $\mathrm{\frac{w_{\lambda,0}}{w_{m,0}}}$, the model displays a decelerating expansion phase similar to the Einstein-de Sitter model, where $q$ approaches 1/2. 
 Furthermore, by substituting Eq.~(\ref{qLCDM}) into  Eq.~(\ref{Reconstruction1}), we can determine the relevant $\mathrm{f(Q)}$  as:
\begin{equation}\label{fLCDM}
  \mathrm{  f(z)=-6H_0^2\left(w_{m,0}(1+z)^3+w_{\lambda,0}\right)-6w_{\lambda,0}H_0^2}\,.
\end{equation}
This outcome aligns with expectations, leading to $\mathrm{f(Q)_{\Lambda CDM}=Q-const.}$, specifically in the $\Lambda$CDM model \cite{Nesseris:2013jea}. Therefore, observations are necessary to constrain the two parameters of the model: $\mathrm{H_0}$ and $\mathrm{\Omega_{m,0}}$. Actually, the CMB and BAO data in the local region, assuming the $\Lambda$CDM model, support a $\mathrm{H_0}=68$ km/s/Mpc and $\mathrm{\Omega_{m,0}}=0.3$, whereas the SNIa and global $\mathrm{H_0}$ data (independent of model) prefer a higher $\mathrm{H_0}=73$ km/s/Mpc and lower $\mathrm{\Omega_{m,0}}=0.26$.
In the context of modified gravity, one can seek an explanation for the accelerated expansion without relying on dark energy (cosmological constant), but significant deviations from the $\Lambda$CDM model are not likely.
\subsection{First application}\label{Sec4.2}
Numerous different ways to parameterize the deceleration parameter have been proposed in studies, however, they typically share a similar structure as given by:
\begin{equation}\label{general-param-q}
{\mathit{    q(z)=q_{0}+q_{1}X(z)}}\,,
\end{equation}
The data sets can determine   $q_{0}$ and $q_{1}$ provided they are real, but  the function $X(z)$ can yield  different representations of the deceleration parameter. Inspired by the parametrization of the dark energy of EoS  presented in \cite{Barboza:2008rh}, a parametrization of the deceleration parameter that is free of divergences was proposed in \cite{AlMamon:2015ali}
\begin{equation}\label{Mod1-Xz}
 \mathrm{   X(z)=\frac{z(1+z)}{1+z^2}}\,.
\end{equation}
\begin{figure*}[t!]
\centering
\subfigure[~$f(Q)$ theory]{\label{fig:Mod1-fz}\includegraphics[scale=0.22]{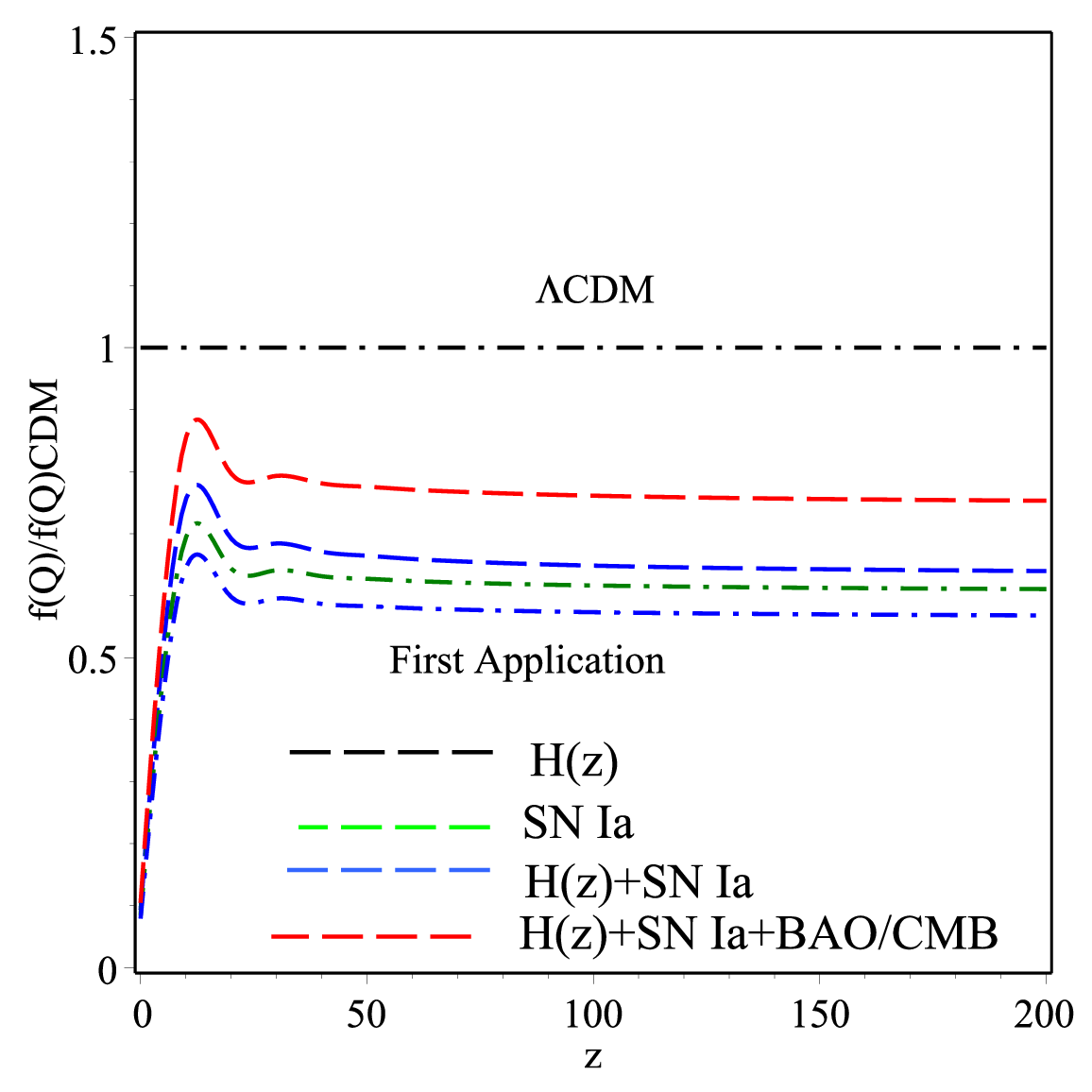}}
\subfigure[~$w_{m}(z)$, $w_{Q}(z)$ behaviors]{\label{fig:Mod1-Om}\includegraphics[scale=0.22]{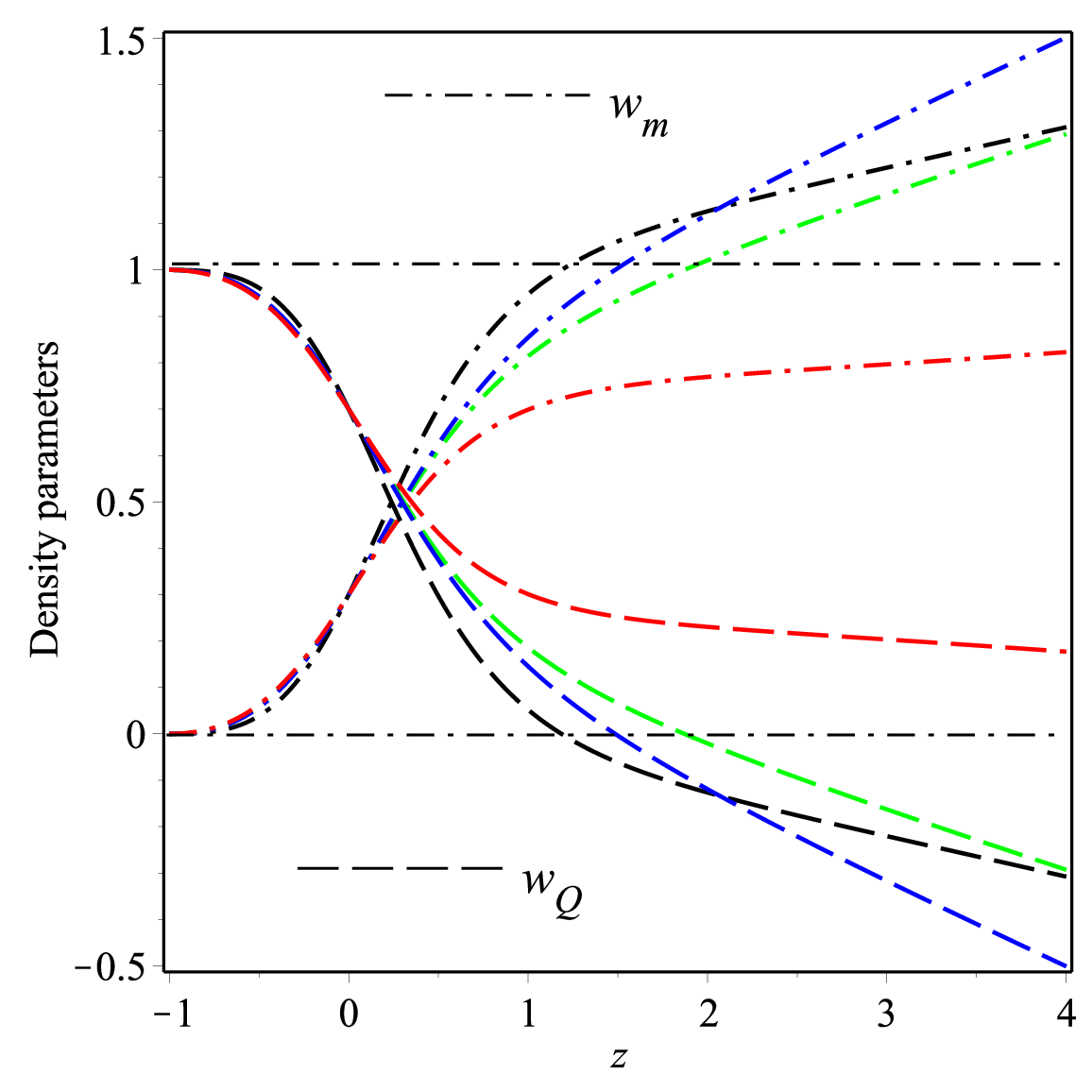}}
\subfigure[~$\omega^{Tot}$]{\label{fig:Mod1-weff}\includegraphics[scale=0.22]{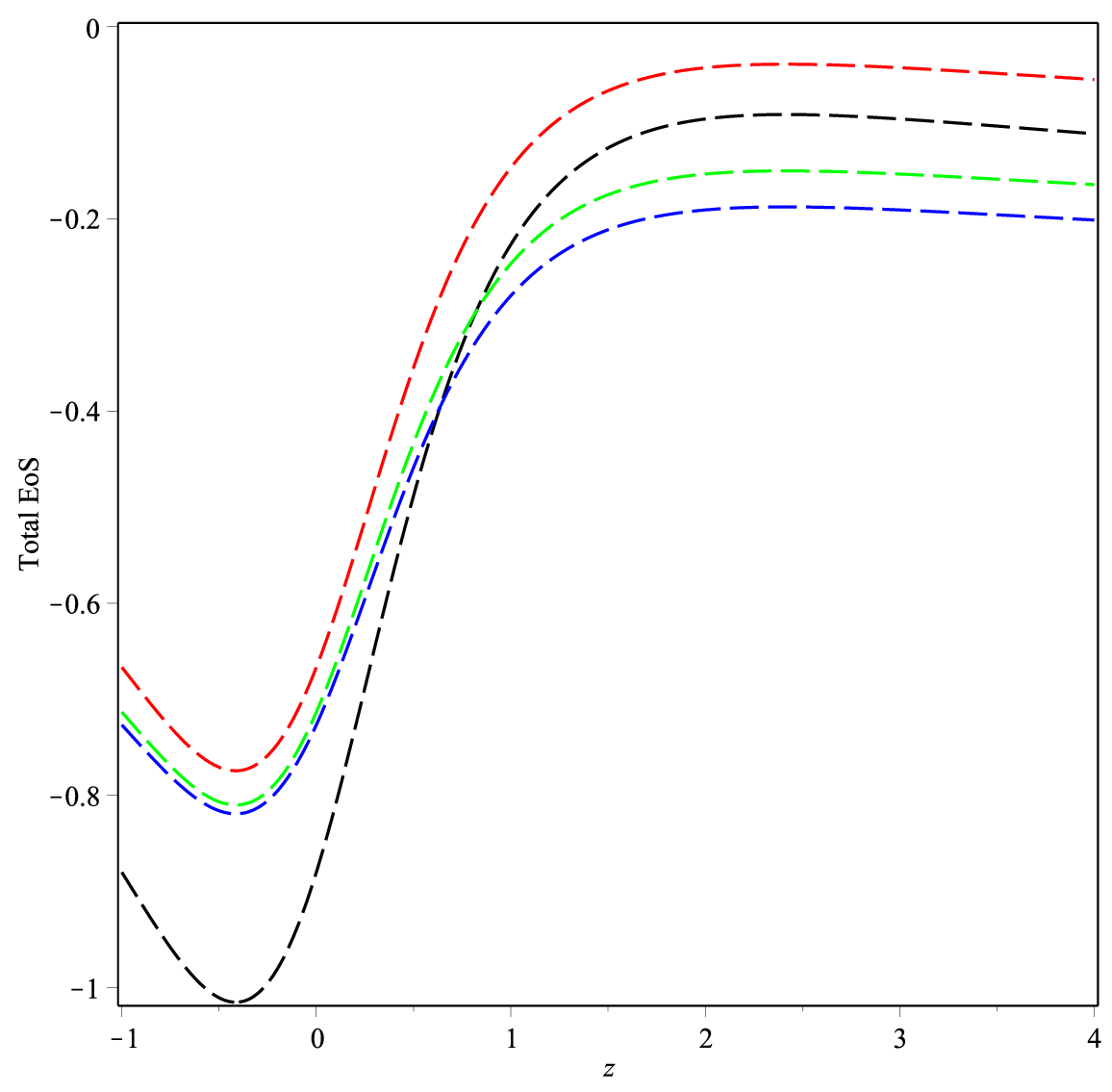}}
\subfigure[~$\omega_{Q}$ behavior]{\label{fig:Mod1-wT}\includegraphics[scale=0.22]{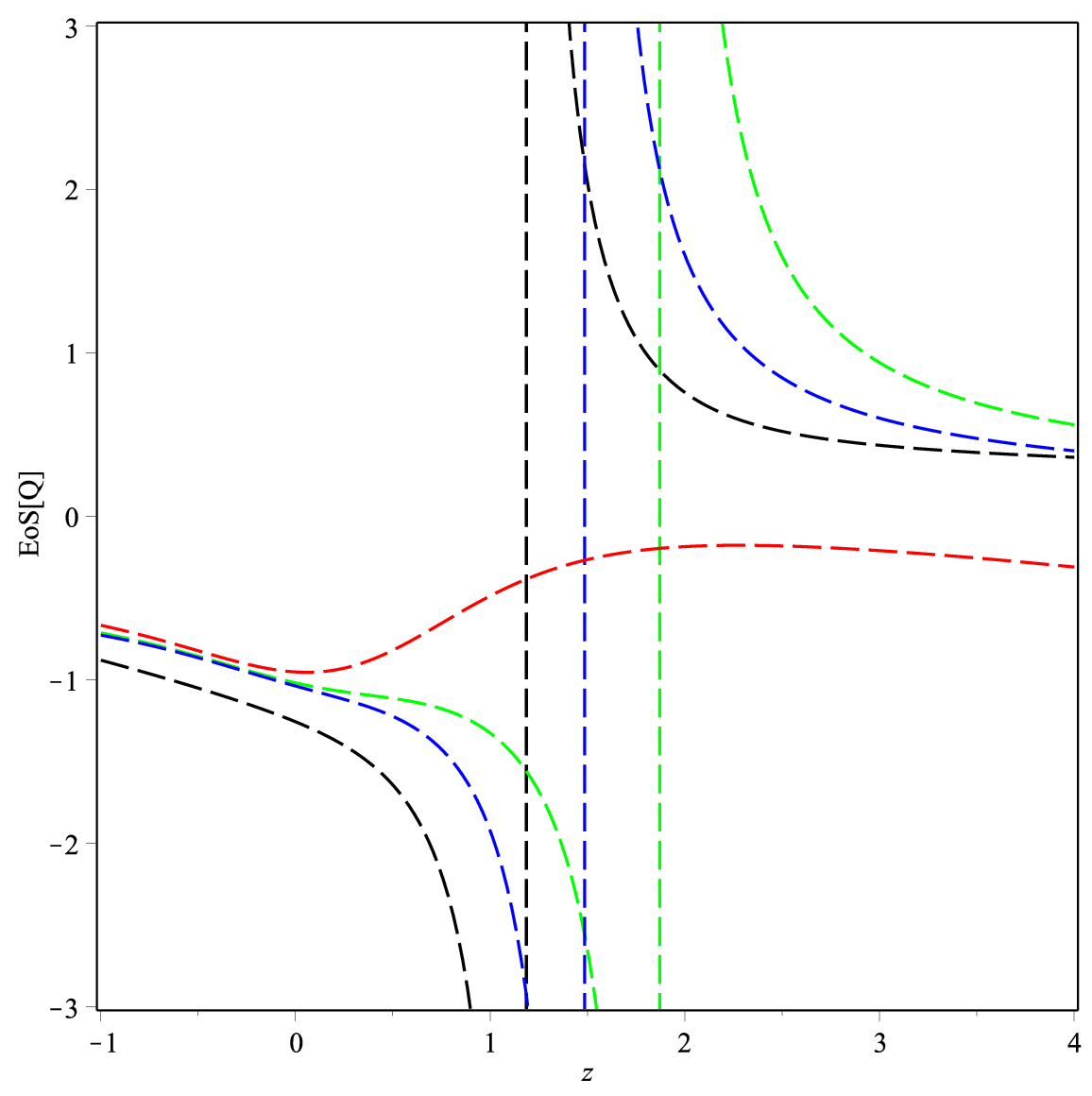}}
%
\caption[figtopcap]{
\subref{fig:Mod1-fz} The behavior of $\mathrm{f(Q(z))}$
\subref{fig:Mod1-Om} The parameter for matter density
\subref{fig:Mod1-weff} Total EoS
\subref{fig:Mod1-wT} The symmetric teleparallel dark energy of EoS.
{ The  and labels used in Fig.~\ref{Fig:Mod1} \subref{fig:Mod1-fz} are the same in Figs.~\ref{Fig:Mod1}\subref{fig:Mod1-Om}, \ref{Fig:Mod1}\subref{fig:Mod1-weff}, and \ref{Fig:Mod1}\subref{fig:Mod1-wT}.
}}\label{Fig:Mod1}
\end{figure*}
At high red-shift $\mathrm {z\gg 1}$, the deceleration parameter $q(z)$ approaches $q_0+q_1$, making it appropriate for analyzing the radiation era. In the late universe with $\mathrm{0\leq z \ll 1}$, $\mathrm{q(z)}$ can be represented as $\mathrm{q(z)=q_0+q_1 z}$.  In addition, it has been proven that $\mathrm q(z)$ remains finite as $\mathrm{z}$ approaches $-1$, making it appropriate for examining the universe's destiny. The parametric form given above is limited for every red-shift $z\in\left[-1, \infty \right)$, making it suitable for depicting the entire cosmic timeline as noted in \cite{AlMamon:2015ali}.  By utilizing the parametric Eqs.~(\ref{Mod1-Xz}) and (\ref{Hubble-deceleration}),  we get:
\begin{equation}\label{Mod1-H}
{\mathit{    H(z)=H_{0}(1+z)^{1+q_{0}}(1+z^2)^{\frac{q_{1}}{2}}}}\,.
\end{equation}
Furthermore, through the utilization of Eq.~(\ref{Reconstruction1}), we derive the form of $\mathrm{f(z)}$  as:
\begin{eqnarray}
  {\mathit{ f(z)=-6w_{m,0}H_0^2(1+z)^{1+q_{0}}(1+z^2)^{\frac{q_{1}}{2}}
\int_0^z {\frac{1+q_{0}+q_{1}\frac{\tilde{z}(1+\tilde{z})}{1+\tilde{z}^2}}{(1+\tilde{z})^{q_{0}-1}(1+\tilde{z}^{2})^{\frac{q_{1}}{2}}}} d\tilde{z}}}\,.\label{Mod1-f(z)}
\end{eqnarray}
Figure \ref{Fig:Mod1}\subref{fig:Mod1-fz} displays the $f\mathrm{(Q)}$ gravity results as a function of red-shift for various $\mathrm {q_0}$ and $\mathrm{q_1}$ parameters. Figure  \ref{Fig:Mod1}\subref{fig:Mod1-fz}  indicate  significant discrepancies between the theory and the $\Lambda$CDM model, which is not commonly accepted.  
\begin{figure*}[t!]
\centering
\subfigure[~$f(Q)$  ]{\label{fig:Mod1A-fz}\includegraphics[scale=0.22]{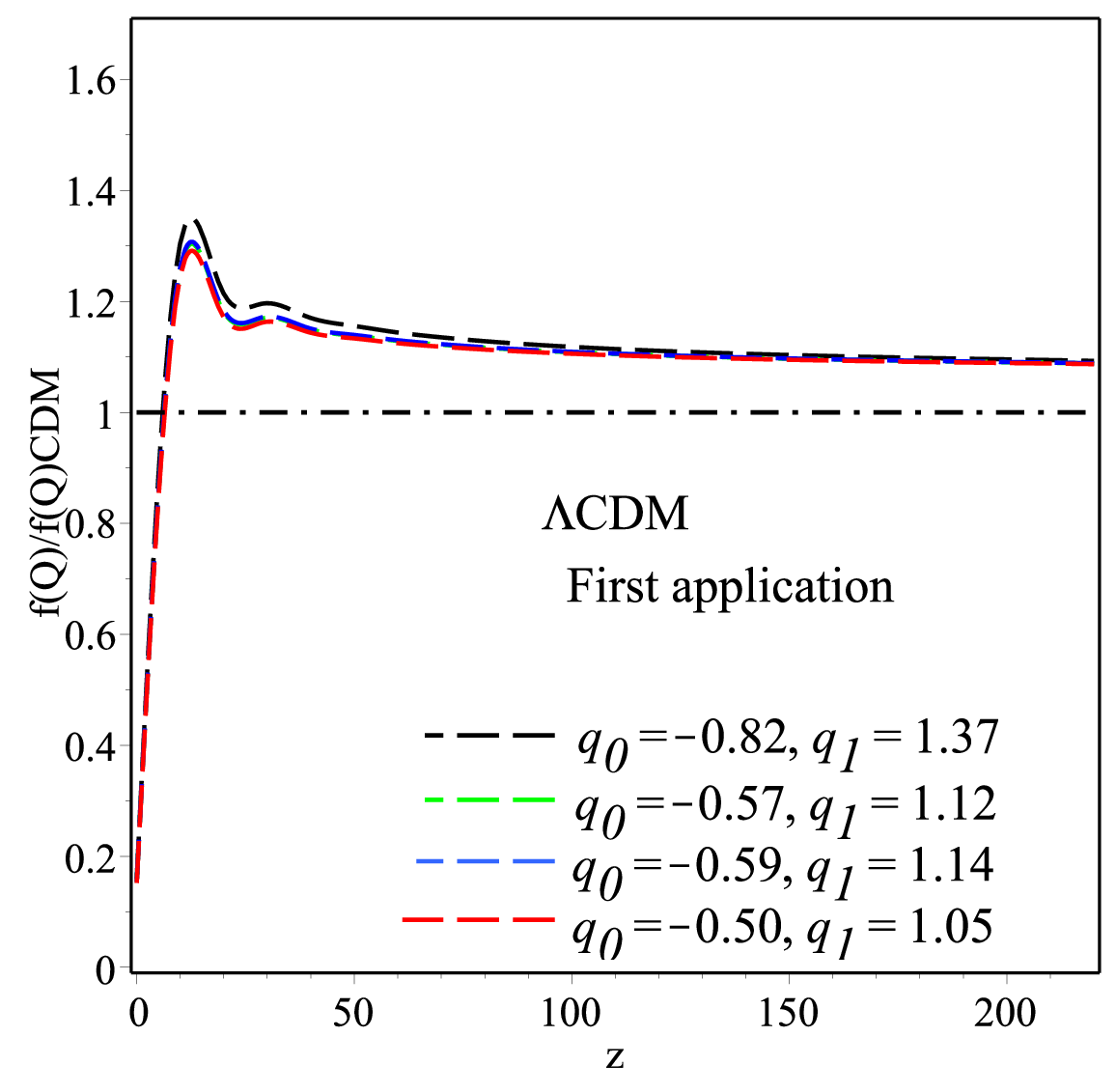}}
\subfigure[~$w_{m}$, $w_{Q}$ ]{\label{fig:Mod1A-Om}\includegraphics[scale=.22]{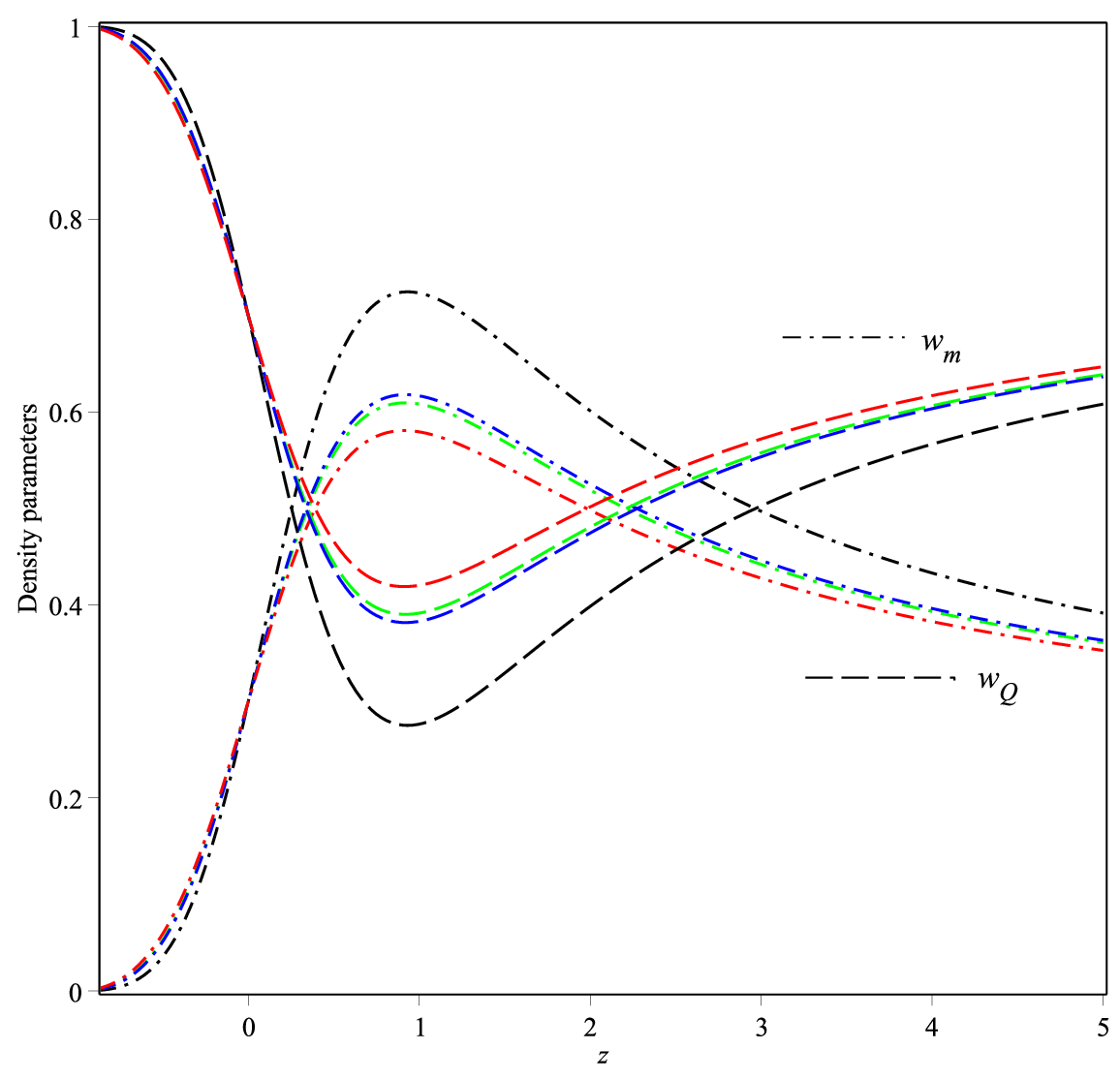}}
\subfigure[~$\omega^{Tot}$ ]{\label{fig:Mod1A-weff}\includegraphics[scale=.22]{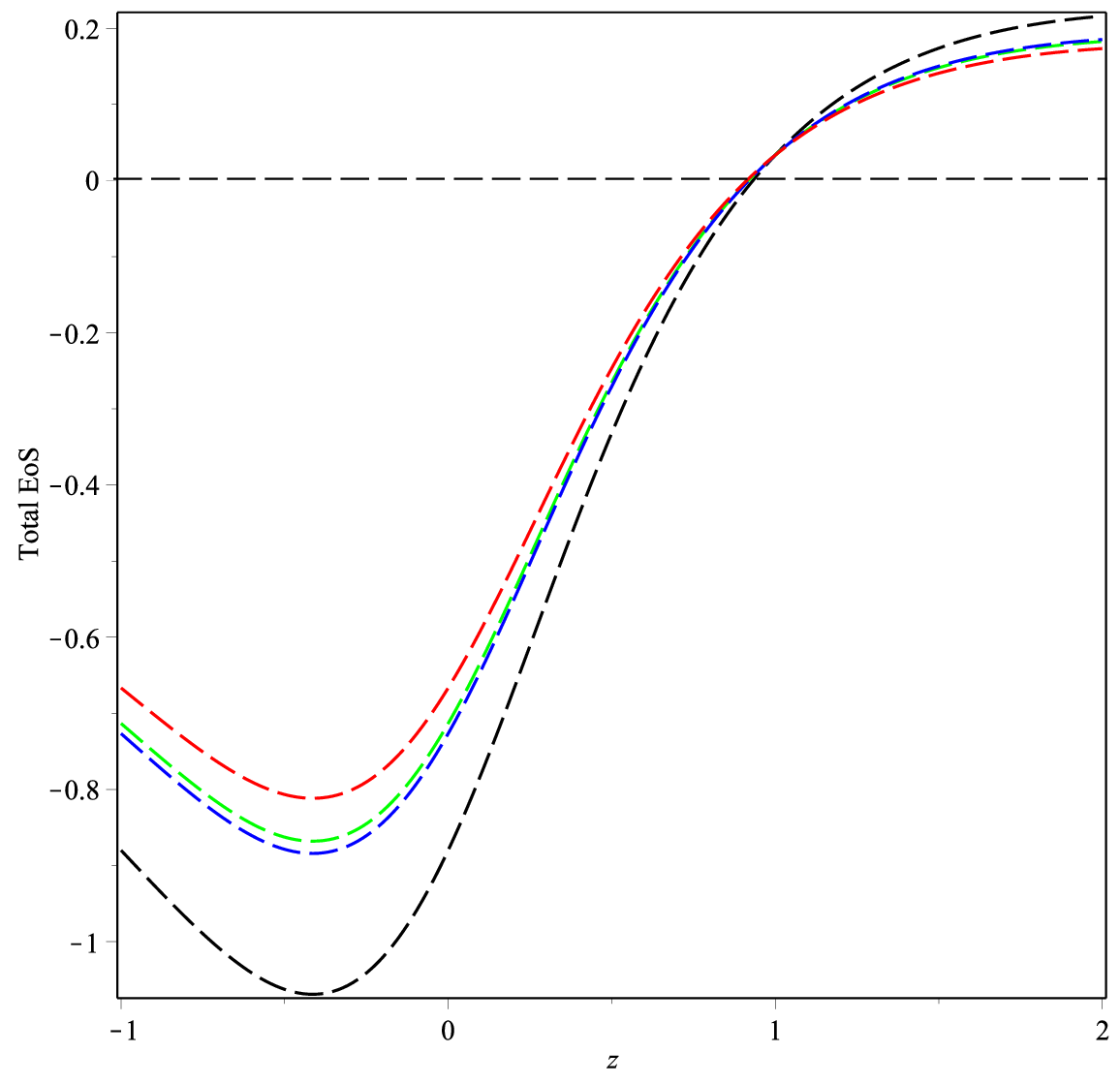}}
\subfigure[~$\omega_{Q}$ ]{\label{fig:Mod1A-wT}\includegraphics[scale=.22]{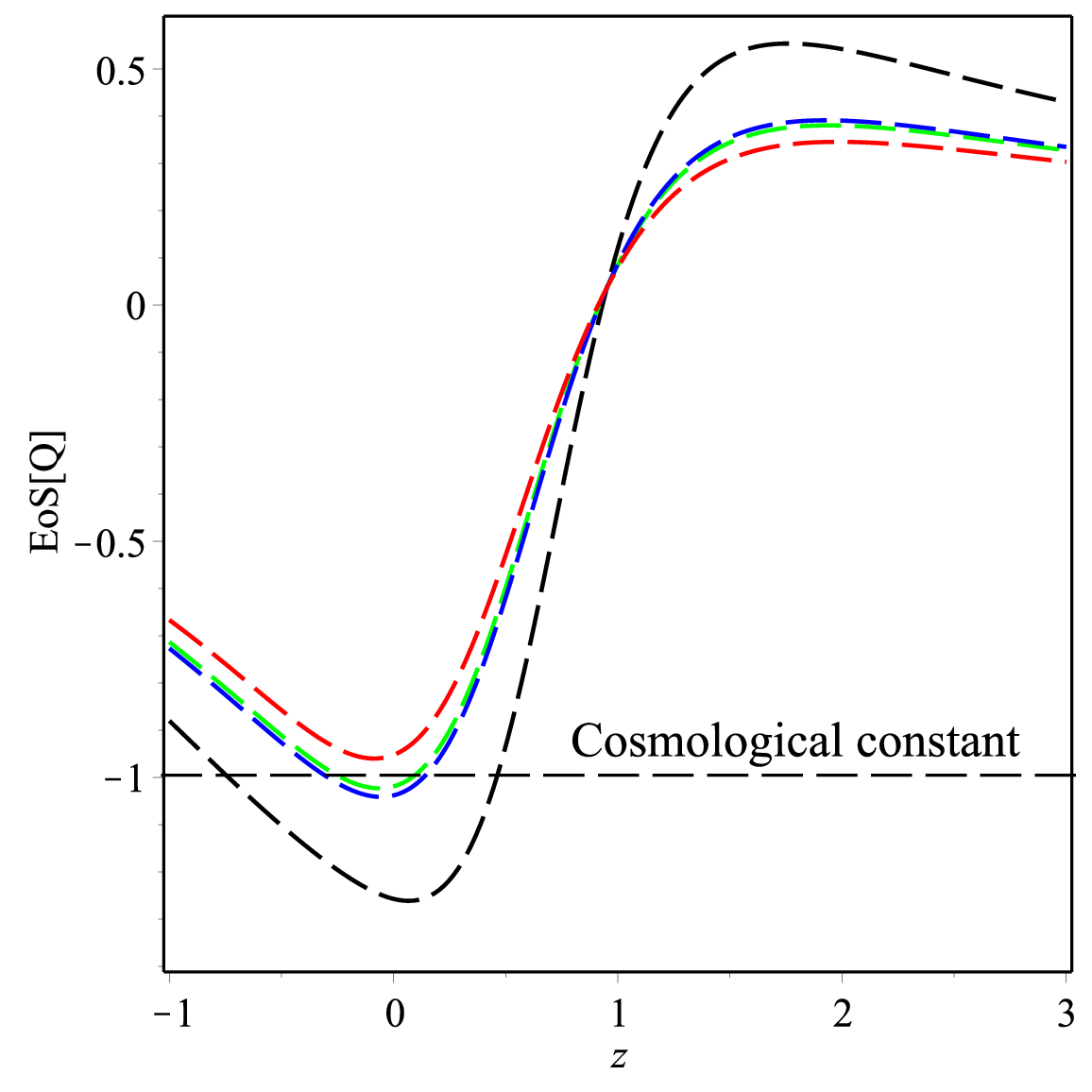}}
%
\caption[figtopcap]{\small{
Fig. \subref{fig:Mod1A-fz} represents the behavior of $\mathrm{f(Q)}$;
\subref{fig:Mod1A-Om} The parameter of matter density;
\subref{fig:Mod1A-weff} Total  EoS;
\subref{fig:Mod1A-wT} The non-metricity EoS. { The $q_0$ and $q_1$ values that are presented in Fig.~\ref{Fig:Mod1A}\subref{fig:Mod1A-fz} are used in Figs.~\ref{Fig:Mod1A}\subref{fig:Mod1A-Om}, \ref{Fig:Mod1A}\subref{fig:Mod1A-weff}, and \ref{Fig:Mod1A}\subref{fig:Mod1A-wT}.}
}}
\label{Fig:Mod1A}
\end{figure*}

Following Eq. (\ref{deceleration}), we have the following expression:
\begin{equation}\label{Mod1-weff}
  {\mathit{  \omega^{Tot}(z)=-1+\frac{2}{3}\frac{(1+q_{0})+q_{1}z+(1+q_{0}+q_{1})z^2}{1+z^2}}}\,.
\end{equation}

\begin{table*}[t!]
\caption{\label{Table1}%
First application's primary outcomes are determined by ($\mathrm{q_0,q_1}$)  presented in \cite{AlMamon:2015ali} and the matter density parameter restriction (\ref{q1-min}). Later on, we maintain the stringent measured values of $\mathrm{q_0}$, but modify $\mathrm{q_1}$ as necessary to satisfy the restriction.}
\begin{ruledtabular}
\begin{tabular*}{\textwidth}{lcccccccc}
\multirow{2}{*}{\textbf{Dataset}}                        & \multirow{2}{*}{$q_0$} && \multirow{2}{*}{$q_1$} & $\mathrm{f(Q)}/\Lambda$CDM  & $w_m(z)\leq 1$          & \textbf{Non-metricity} & 
 \multirow{2}{*}{\textbf{Viability}} \\
                                        &       &&       & \textbf{conformity} & \textbf{limit} & \textbf{EoS}, $\mathrm{w_Q}$
                                           &  \\
\hline
H(z)                                    &       $-0.82$      &&      $0.98$      & not                          & violated ($z\sim 1.31$)                             & diverges
 ($z\sim 1.31$) &  not                 \\
SNIa                                    &     $-0.57$        &&       $0.70$      & not                          & broken ($z\sim 1.29$)                             & diverges$^\textrm{\ref{footnote:1a}}$ ($z\sim1.29$)         & not                 \\
H(z)+SNIa                               &       $-0.59$      &&      $0.67$       & not                         & violated ($z\sim1.5$)                             & diverges$^\textrm{\ref{footnote:1a}}$ ($z\sim1.5$)       & not                 \\
H(z)+SNIa+BAO/CMB                       &        $-0.50$     &&       $0.78$      & not                          & violated ($z\sim 0.82$)                            & finite ($z\sim 0.82$)    &not                 \\
\hline
\textbf{Using constraint (\ref{q1-min})}    & \textbf{}   && \textbf{}   & \textbf{}                   & \textbf{}                      & \textbf{}        & \textbf{}          \\
\hline
H(z)              &      $-0.82$       &&     $1.37$        & approximately                        & fulfilled
  & finite    &not
        \\
SNIa              &       $-0.57$       &&     $1.12$        & approximately                         & fulfilled
& finite     &not
\\
H(z)+SNIa         &        $-0.59$      &&     $1.14$        & approximately                       & fulfilled
              &  finite  & not
                   \\
H(z)+SNIa+BAO/CMB &       $-0.50$      &&       $1.05$      & approximately                        & fulfilled
                      & finite    & not
\end{tabular*}
\end{ruledtabular}
\end{table*}

By utilizing the parameterizations given in equation (\ref{Mod1-Xz}), the matter density parameter in equation (\ref{matter-density-parameter}) can be expressed as:
\begin{equation}\label{Mod1-Omega_m}
{\mathit{    w_{m}(z)=w_{m,0}(1+z)^{1-2q_{0}}(1+z^2)^{-q_{1}}}}\,.
\end{equation}
Based on the given values of the model parameters, Figure \ref{Fig:Mod1}\subref{fig:Mod1-Om}  displays the changes in ${\mathit {w_{m}(z)}}$ and  ${\mathit {w_{Q}}}$ where the parameters $q_0$ and $q_1$ are listed in \cite{AlMamon:2015ali}. The plots reveal that the matter density parameter reaches the value of one between red-shift $1\lesssim z \lesssim2$ for  $\mathrm{H(z)}$, SNIa, and $\mathrm{H(z)}$+SNIa datasets, but for  $\mathrm{H(z)}$+SNIa+BAO/CMB, it does not  happen.   Based on the analysis of the $\mathrm{f(Q)}$ gravity found in Eq.~(\ref{Mod1-f(z)}), it is not surprising that the parametrization (\ref{Mod1-Xz}) does not result in a  standard CDM that aligns with the thermal history.

 Setting $\mathrm{\omega^{Tot}(z)=-1/3}$ yields the red-shift $\mathrm{z}$ of transition based on  $\mathrm{q_0}$ and $\mathrm{q_1}$ given in \cite{AlMamon:2015ali}.  In  $\mathrm{H(z)}$, SNIa, $\mathrm{H(z)}$+SNIa datasets, and \\$\mathrm{H(z)}$+SNIa+BAO/CMB, this yields $\mathrm{z_\backsimeq 0.71}$, $0.72$, $0.8$, and $0.54$, respectively. Fig. illustrates the evolution of $\omega^{Tot}(z)$. \ref{Fig:Mod1}\subref{fig:Mod1-weff}.  While the graphs display the transition red-shift in line with observations, they do not align with the standard CDM behaviour (i.e $\mathrm{\omega^{Tot}(z)=0}$) at high redshifts. In this way, the model was ineffective at the earlier stages (huge redshifts).  We additionally evaluate the parametric expression (\ref{Mod1-Xz}) associated with the symmetrical teleparallel EoS parameter:
\begin{equation}\label{Mod1-wDE}
 {\mathit{   \omega_{Q}(z)=\frac{2(1+z)^{2q_{0}}\left[(q_{0}-\frac{1}{2})+\frac{q_{1}z(1+z)}{1+z^2}\right]}
    {3\left[(1+z)^{2q_{0}}-\frac{w_{0,m}(1+z)}{(1+z^2)^{q_{1}}}\right]}}}.
\end{equation}

The rest of this application demonstrates that feasible cosmic development can be achieved by constraining the model parameters. If the model parameters  predicted values match their measured values, then the assumed parametrization may accurately depict cosmic history.  The matter density parameter actually needs to get closer to a maximum value of $\mathrm{w_{m}(z)=1}$ as $\mathrm{z}$ gets closer to infinity. Such requirement is helpful in imposing an additional limitation on the independent variables $\mathrm{q_0}$ and $\mathrm{q_1}$.  The matter density parameter (\ref{Mod1-Omega_m}) is expressed asymptotically up to the second order of the red-shift as follows for more clarification as:
\[\mathrm{\widetilde{w}_{m}(z)\thickapprox w_{m,0} \left(\frac{1}{z}\right)^{2(q_{0}+q_{1})}{\mathit (1+z-2q_0)+O(1/z^2)}}\,.\]
In order  to be feasible, $\mathrm{\widetilde{w}_{m}(z)}$ must be less than or equal to 1; alternatively, the framework of symmetric teleparallel gravity would be required to allow the density parameter to be less than zero which restricts $\mathrm{q_{1}}$ to a lower limit.
\begin{equation}\label{q1-min}
\mathrm{ q_{1}\geq \lim_{z\to \infty}\frac{\mathit{\ln \left[z^{-2q_{0}}(1+z-2q_{0})w_{0,m}\right]}{\ln z^{2}} = \frac{1}{2}-q_{0}}}\,.
\end{equation}
Equation (\ref{q1-min}) limits the selection of model parameters to have:
\begin{align}\label{con}  {\mathit {q_{0}+q_{1}\geq 0.5}}\,.\end{align}
Historically, $\mathit{    w_{m}(z)}$  would surpass 1 at a certain redshift if the condition $\mathrm{q_{0}+q_{1}} < 0.5$ was met. The transition of $\mathrm{w_{m}(z)}$, the equation of state parameter for matter, occurs closer to the unit boundary at a higher redshift $\mathrm{z}$, particularly as $\mathrm{q_{0}+q_{1}}$ approaches 0.5 from below, denoted as $\mathrm{q_{0}+q_{1} \to 0.5^{-}}$. This indicates that the timing of the transition is sensitive to the sum of the current deceleration parameter $\mathrm{q_{0}}$ and an additional parameter $\mathrm{q_{1}}$, with transitions happening earlier (at larger $\mathrm{z}$) as this sum gets closer to 0.5.
\subsection{Second application}\label{Sec4.3}
In this application, we are going to investigate an alternative parametrization $\mathrm{q(z)}$ as  \cite{Mamon:2016dlv}: \begin{equation}\label{Mod2-Xz}
 \mathrm{   X(z)=\frac{\ln{(\alpha+z)}}{(1+z)}-\ln{\alpha}\,, \qquad \alpha>1}\,.
\end{equation}
\begin{figure*}[t!]
\centering
\subfigure[~$\mathrm {f(Q)}$ behavior]{\label{fig:Mod2-fz}\includegraphics[scale=0.22]{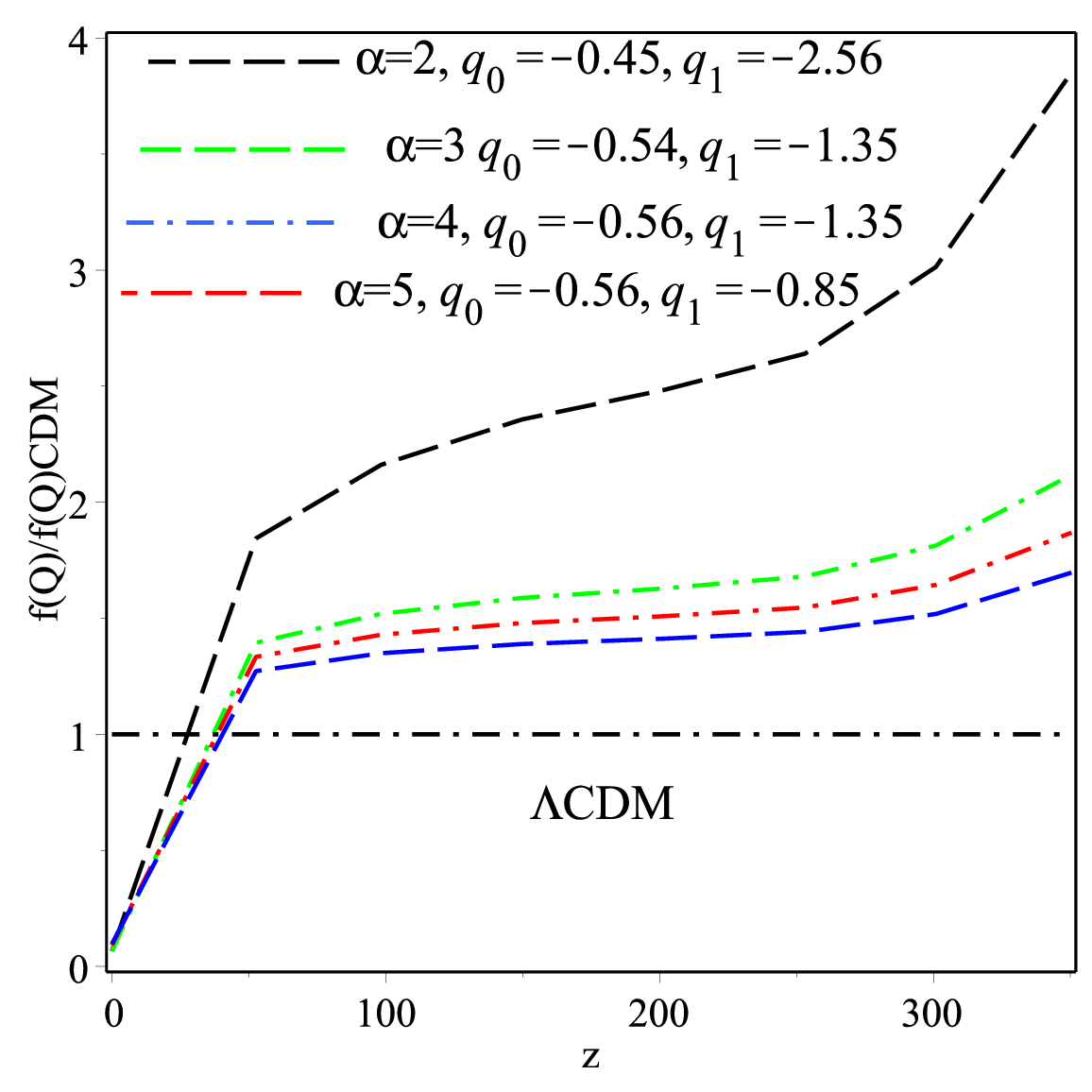}}
\subfigure[~$\mathrm{w_{m}(z)}$, $\mathrm{w_{Q}(z)}$ behavior]{\label{fig:Mod2-Om}\includegraphics[scale=.22]{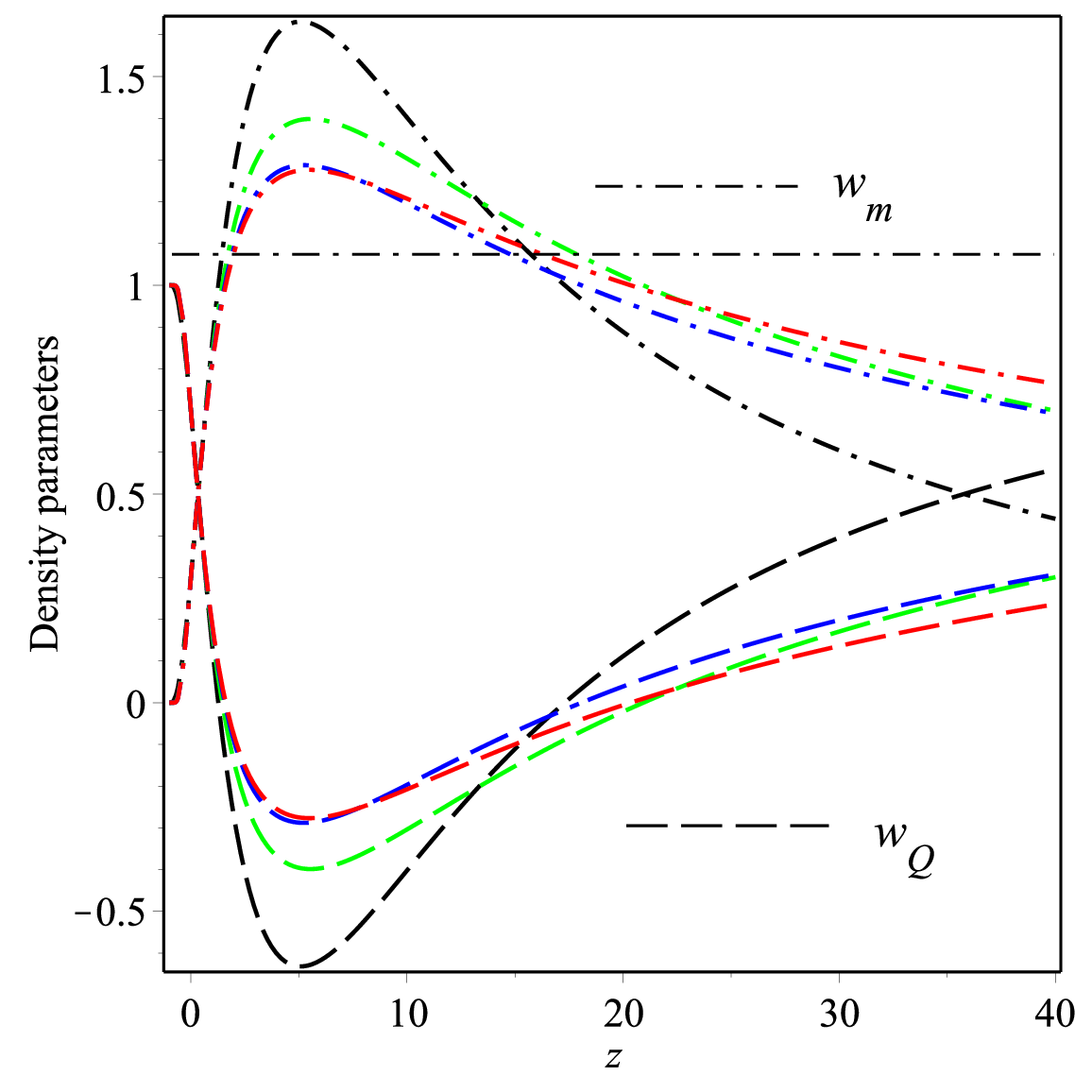}}
\subfigure[~$\omega^{Tot}$ behavior]{\label{fig:Mod2-weff}\includegraphics[scale=.22]{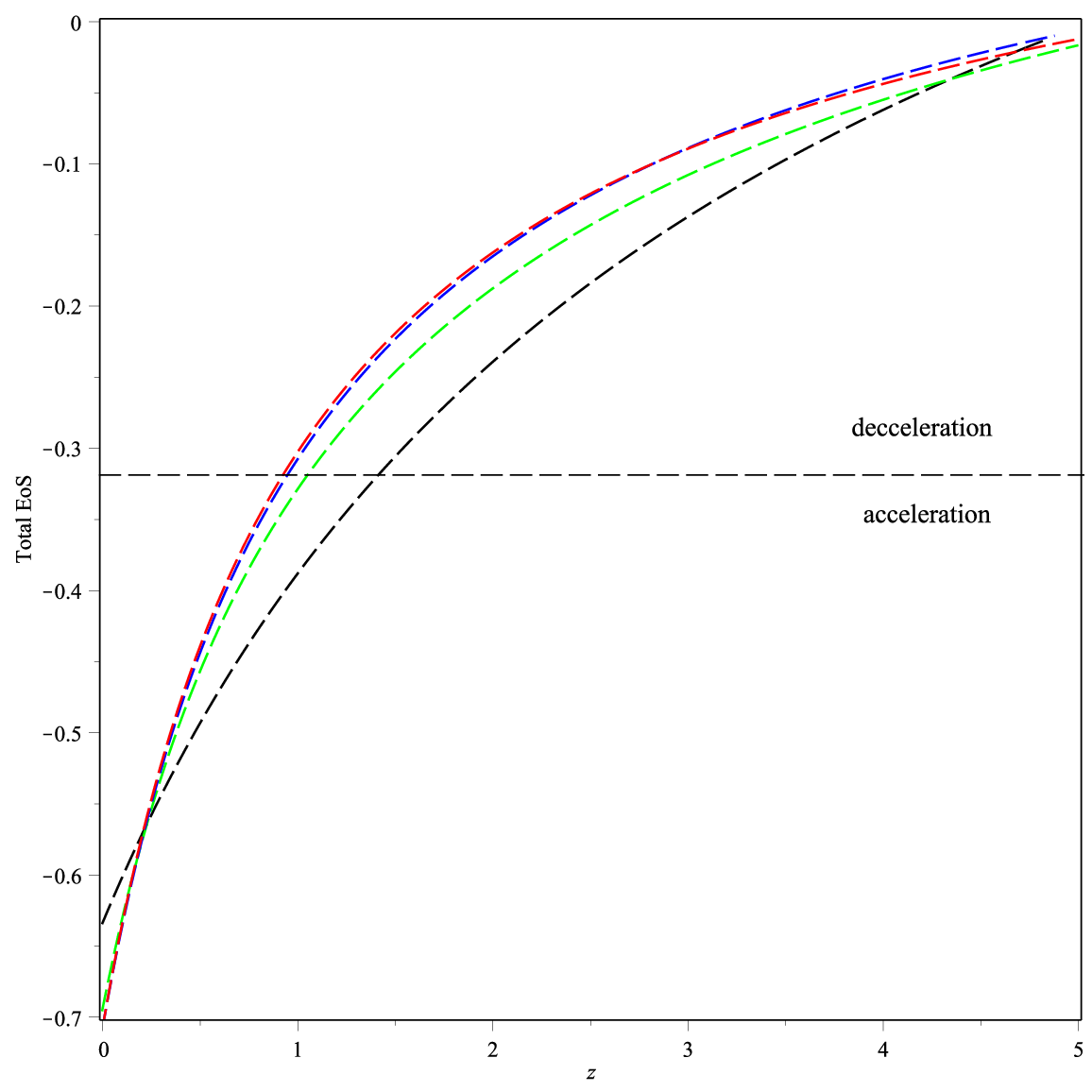}}
\subfigure[~$\mathrm{\omega_{Q}}$ behavior]{\label{fig:Mod2-wT}\includegraphics[scale=.22]{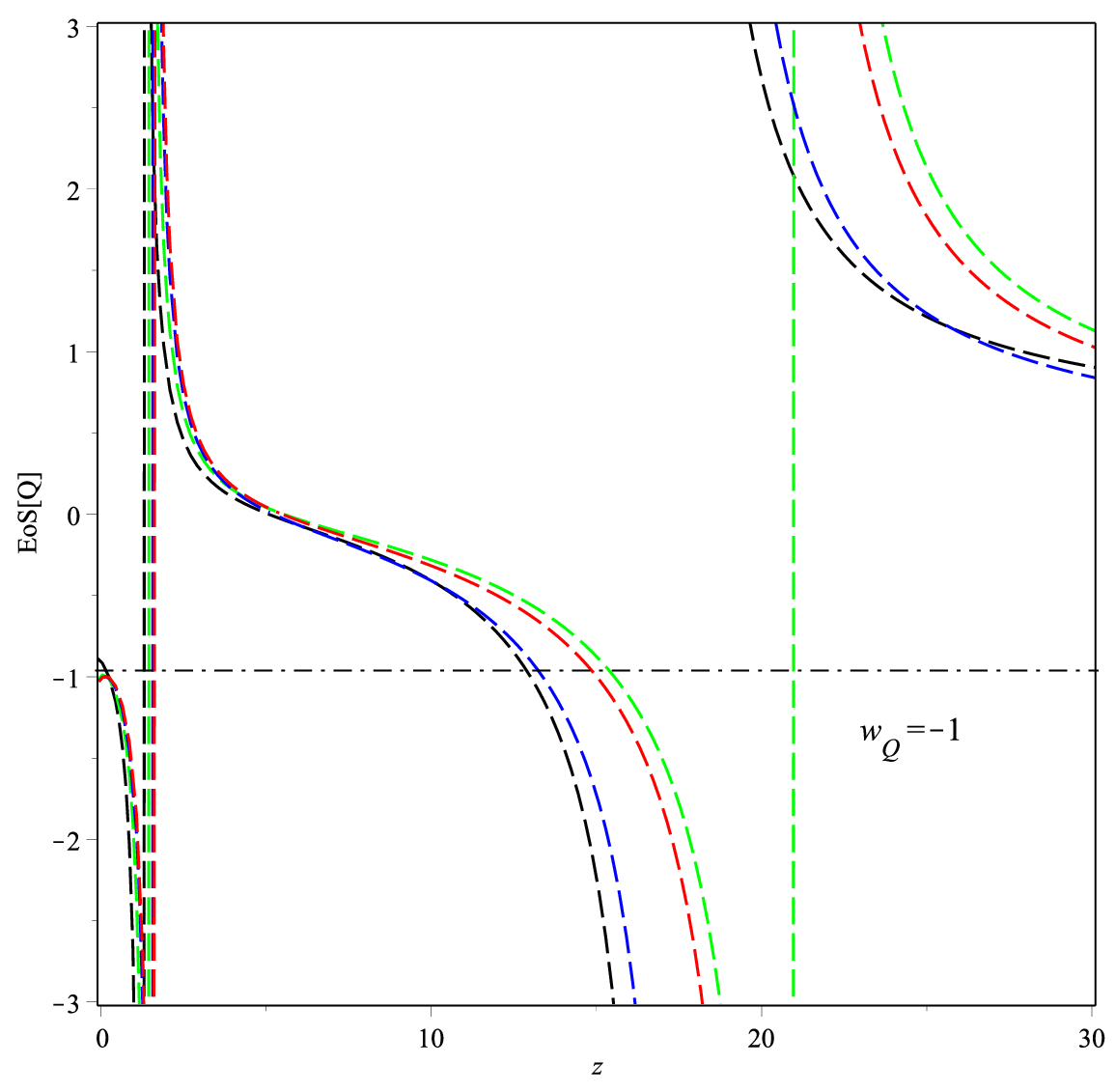}}
\caption[figtopcap]{\small{
%
\subref{fig:Mod2-fz} The behavior of $\mathrm{f(Q(z))}$
\subref{fig:Mod2-Om} Matter density parameter
\subref{fig:Mod2-weff} The total EoS
\subref{fig:Mod2-wT} The EoS for for symmetric teleparallel dark energy. { The  $q_0$ and $q_1$  numerical values presented in Fig.~\ref{Fig:Mod2}\subref{fig:Mod2-fz} are used in Figs. \ref{Fig:Mod2}\subref{fig:Mod2-Om}, \ref{Fig:Mod2}\subref{fig:Mod2-weff}, and \ref{Fig:Mod2}\subref{fig:Mod2-wT}}.
}}
\label{Fig:Mod2}
\end{figure*}
By utilizing the parametrization mentioned above, the deceleration parameter (\ref{general-param-q}) was tested against observational data sets, 
 Furthermore, it has been demonstrated that at $\textit{z=0}$, the parameterizations lead to  $\mathrm{q=q_0}$. Moreover, the condition $\mathrm{q_1=\frac{2q_0-1}{2\ln{\alpha}}}$ applied at high redshifts enables the realization of the matter-dominated era, where ${\textit q=\frac{1}{2}}$.  Through the use of the parametric form given by equations (\ref{Mod2-Xz}) and (\ref{Hubble-deceleration}), the Hubble-red-shift relation can be expressed.
\begin{figure*}[t!]
\centering
\subfigure[~$\mathrm{f(Q)}$ theory]{\label{fig:Mod2A-fz}\includegraphics[scale=0.22]{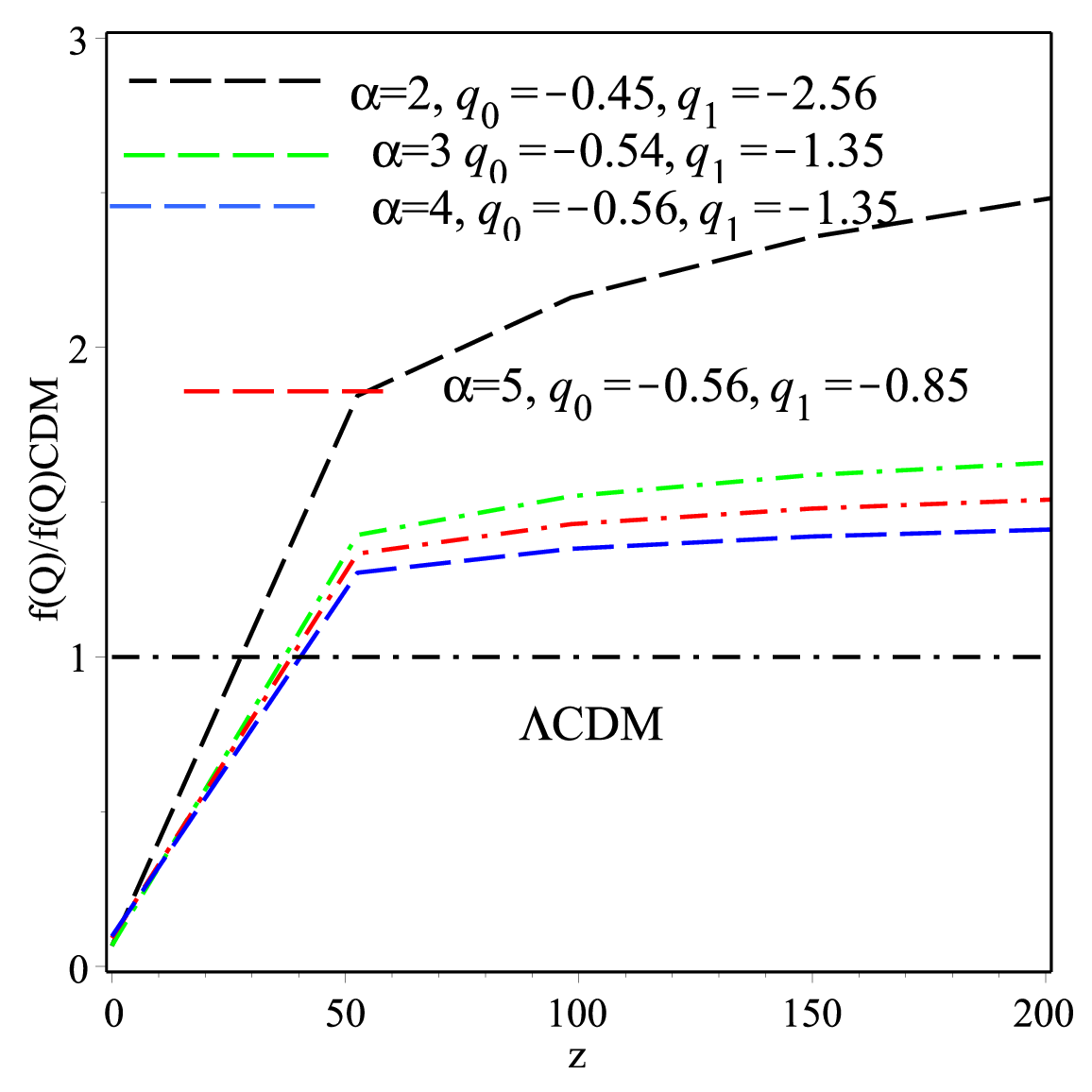}}
\subfigure[~$\mathrm{w_{m}}$, $\mathrm{w_{Q}}$ behavior ]{\label{fig:Mod2A-Om}\includegraphics[scale=.22]{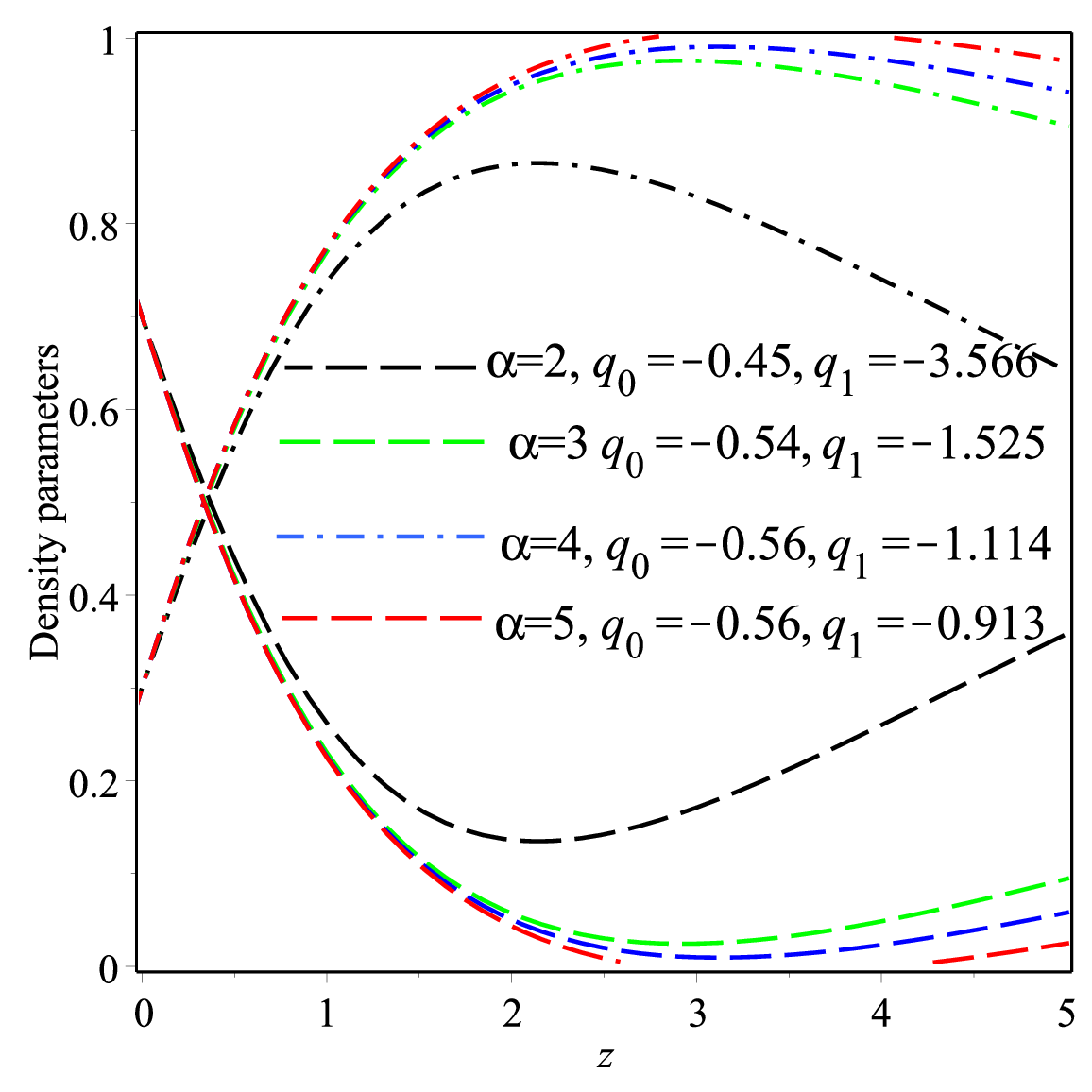}}
\subfigure[~$\mathrm{\omega^{Tot}}$ behavior]{\label{fig:Mod2A-weff}\includegraphics[scale=.22]{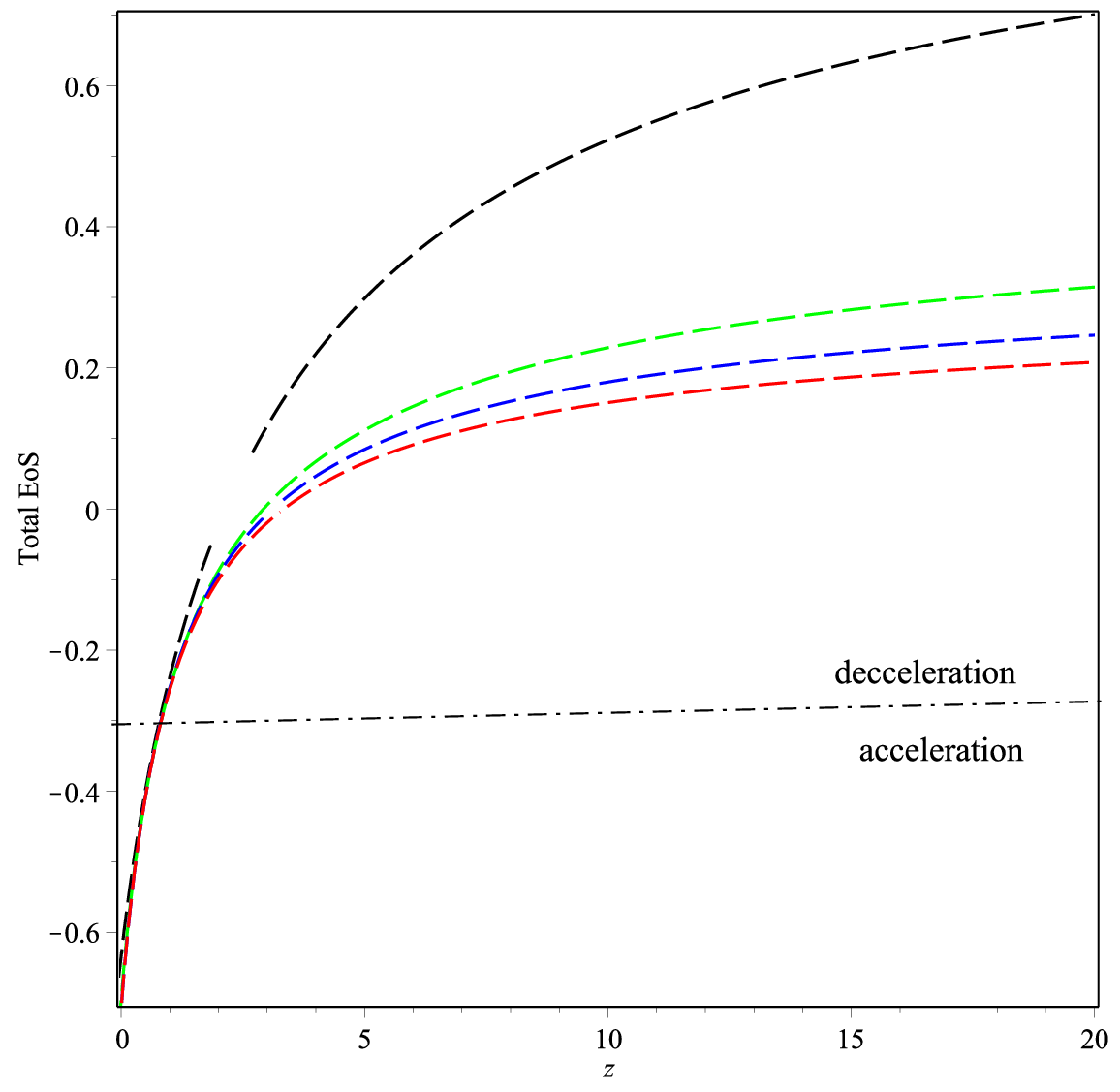}}
\subfigure[~$\mathrm{\omega_{Q}}$ behavior]{\label{fig:Mod2A-wT}\includegraphics[scale=.22]{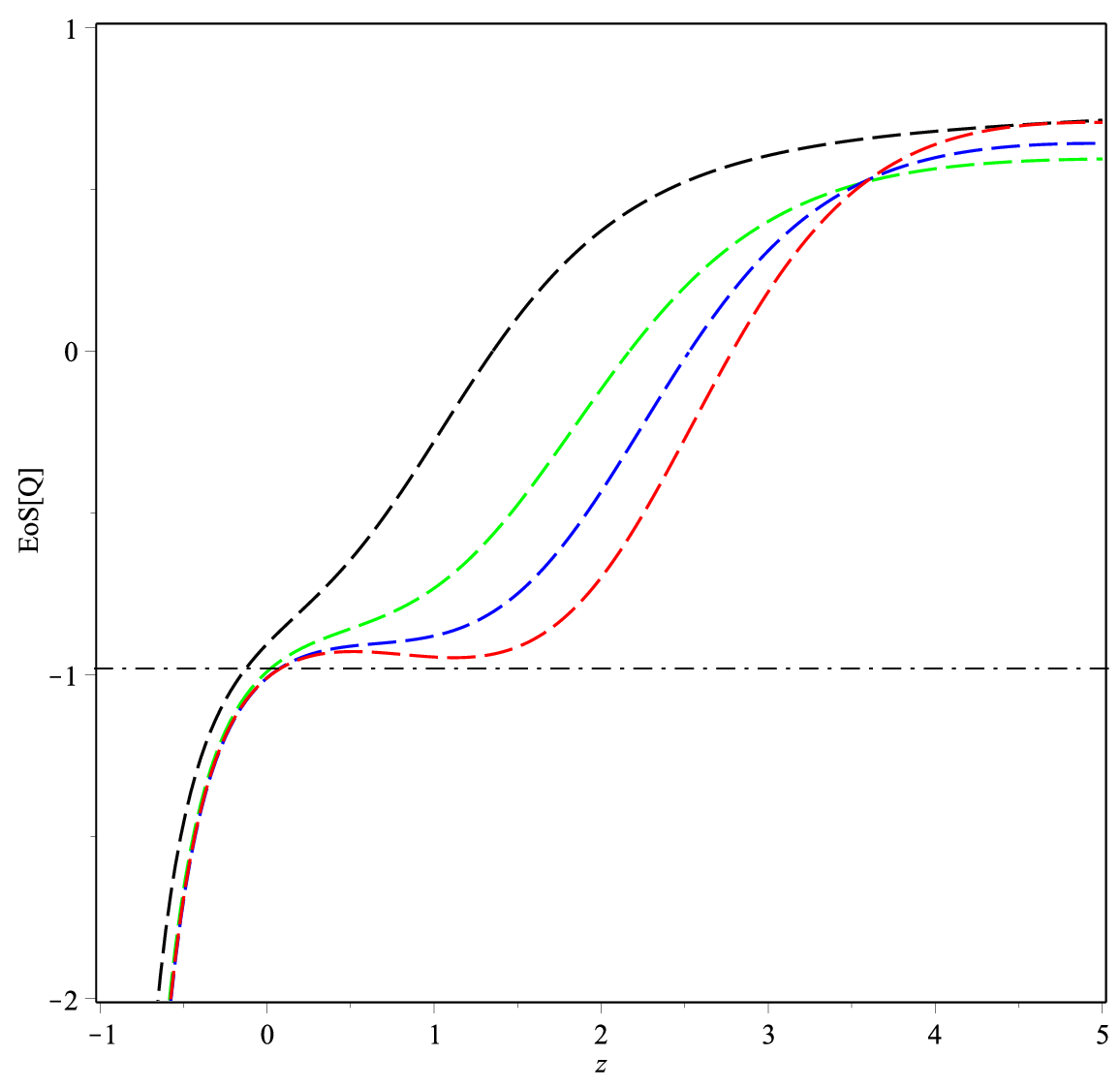}}
\caption[figtopcap]{\small{The values of second application  where  ($\mathrm{q_{0},~q_{1}}$) are determined based on the restriction  (\ref{q2-min}) for the optimal fit. The value of  $\mathrm{q_0}$ remains unchanged from the measurement in \cite{Mamon:2016dlv}, as it aligns with contemporary observations of other cosmological parameters.  Nevertheless, $\mathrm{q_1}$ is adjusted in order to satisfy Eq.~(\ref{q2-min})  shown in the graphs,
\subref{fig:Mod2A-fz} The $\mathrm{f(Q)}$
\subref{fig:Mod2A-Om} The density parameter of matter 
\subref{fig:Mod2A-weff} The total EoS 
\subref{fig:Mod2A-wT} the dark energy EoS symmetrical teleparallel 
{ The numerical values of $q_0$ and $q_1$ presented in Fig.~\ref{Fig:Mod2A}\subref{fig:Mod2A-Om} are used in Figs.~\ref{Fig:Mod2A}\subref{fig:Mod2A-weff}, and \ref{Fig:Mod2A}\subref{fig:Mod2A-wT}.}
}}
\label{Fig:Mod2A}
\end{figure*}
\begin{equation}\label{Mod2-H}
    {\mathit{H(z)=H_{0}\alpha^\xi (1+z)^\beta(\alpha+z)^{-\frac{(\alpha+z)\xi}{(1+z)\alpha}}}},
\end{equation}
with $\xi$ being  determined by $\frac{q_1 \alpha}{\alpha-1}$ and $\beta$ is calculated as $1+q_0+\frac{\xi}{\alpha}-q_1\ln{\alpha}$. Moreover, through the utilization of Eq.~(\ref{Reconstruction1}), we derive the form of $\mathrm{f(z)}$
as:
\begin{eqnarray} \label{Mod2-f(z)}
{ \mathit{f(z)=-6\Omega_{m,0}H_0^2\frac{(1+z)^\beta}{(\alpha+z)^{\frac{(\alpha+z)\xi}{(1+z)\alpha}}}\int_0^z\frac{(1+\tilde{z})^{2-\beta}\left(1+q_0+q_1 \frac{\ln{(\alpha+\tilde{z})}}{(1+\tilde{z})}-\ln{\alpha}\right)}{(\alpha+\tilde{z})^{-\frac{(\alpha+\tilde{z})\xi}{(1+\tilde{z})\alpha}}}~d\tilde{z}}}\,.
\end{eqnarray}
Figure \ref{Fig:Mod2}\subref{fig:Mod2-fz}, shows the $\mathrm{f(Q)}$   against the red-shift for various measurements of $\mathrm{q_0}$ and $\mathrm{q_1}$.  Just like the first application, the charts demonstrate significant differences between the theory and the $\Lambda$CDM model, that  is not frequently chosen. Changes in cosmic parameters, such the matter density parameter, must demonstrate this.

Next we will show how Eq. (\ref{deceleration}) is formulated in terms of $z$ for the second application as:
\begin{equation}\label{Mod2-weff}
{\mathit{    \omega^{Tot}(z)=-\frac{1}{3}+\frac{2}{3}\left[q_0+q_1\left(\frac{\ln{(\alpha+z)}}{(1+z)}-\ln{\alpha}\right)\right]}}\,.
\end{equation}
The transition red-shift $\mathrm{z}$ can be found by setting $\mathrm {\omega^{Tot}(z)=-1/3}$ \cite{Mamon:2016dlv}.  According to SNIa+BAO/CMB datasets, the transition red-shift is approximately $1.32$, $0.98$, $0.88$, and $0.86$ for different values of $\alpha$ ranging from $2$ to $5$.  By utilizing Eq.~ (\ref{Mod2-Xz}), then  Eq.~(\ref{matter-density-parameter}) can be expressed as:
\begin{equation}\label{Mod2-Omega_m}
 \mathrm{w_{m}(z)=\frac{w_{m,0}}{\alpha^{2\xi}} (1+z)^{3-2\beta} (\alpha+z)^{2\frac{(\alpha+z)\xi}{(1+z)\alpha}}}\,.
\end{equation}
Fig. \ref{Fig:Mod2}\subref{fig:Mod2-Om} shows how the matter and symmetric teleparallel density parameters evolve based on the specific parameter values $\mathrm{q_0}$ and $\mathrm{q_1}$, derived and presented in the study mentioned as \cite{Mamon:2016dlv}, have been recalculated. The visual representations clearly indicate that   $\mathrm{w_{m}(z)}$ ranges from 0 to 1 during periods of high redshift.
  Yet, the density decreases once more at redshifts as it passes the unit boundary line.
  According to the evaluation of the derived $\mathrm{f(Q)}$ gravity, as shown in Eq. (\ref{Mod2-f(z)}), the parameterizations (\ref{Mod2-Xz}) does not give rise to a period dominated by matter that aligns with the thermal history.

The plot in Fig. \ref{Fig:Mod2}\subref{fig:Mod2-weff} shows the variation of $\mathrm {\omega^{Tot}(z)}$ over time. Despite the plots displaying transition red-shift in line with observations, they do not align with the standard CDM behavior ( $\mathrm {\omega^{Tot}(z)=0}$) at high red-shift.  This reaffirms the lack of effectiveness of the model during the initial stages  i.e., at high redshifts.

Additionally, we assess the symmetric teleparallel parameter linked to the parametric expression (\ref{Mod2-Xz}) as:
\begin{equation}\label{Mod2-wDE}
  {\mathit{  \omega_{Q}(z)=\frac{-1+2q_0+2q_1\left[\frac{\ln(\alpha+z)}{1+z}-\ln{\alpha}\right]}{3-3w_{m,0}\alpha^\xi(1+z)^{3-2s}(\alpha+z)^{\frac{2\xi(\alpha+z)}
  {\alpha(1+z)}}}}}\,.
\end{equation}
Based on  $\mathrm{q_0}$ and ${\mathrm q_1}$ values described in \cite{Mamon:2016dlv}, an illustration of the behavior of $\mathrm{\omega_Q(z)}$ is shown in plot \ref{Fig:Mod2}\subref{fig:Mod2-wT}. The depicts reveals that the symmetrical teleparallel  EoS   transitions to a phantom-like state at lower redshifts. We observe that a rapid change in phase happens at redshifts approximately $\mathrm{z}=15.78$, $17.96$, $14.89$, and $16.41$ for the parameters $\mathrm{\alpha}$ ranging from 2 to 5, when $\mathrm{\omega_{Q}}$ approaches infinity in opposite directions.
According to the model, a steady shift towards a state resembling quintessence will eventually occur over the phantom division line. Notably, as shown in Figs., phase changes in symmetric teleparallel theory  are associated with the crossing of the unit boundary line by $\mathrm{w_m(z)}$, the matter density parameter.

 In particular, \ref{Fig:Mod2}\subref{fig:Mod2-Om} and \ref{Fig:Mod2}\subref{fig:Mod2-wT}.

Next, we demonstrate how possible cosmic development can be produced by limiting the model parameter. If the model's predicted parameter values match the measured values, then the chosen parameterizations could accurately describe the history of the universe. In practice, when $\mathrm z$ approaches infinity,  $\mathrm{w_{m}(z)}$ must approach a maximum value $\mathrm w_m(z)=1$.  This requirement is beneficial for imposing an additional restriction on the independent variables.
 For additional clarity, we expand the matter density parameter in an asymptotic series (\ref{Mod2-Omega_m}) up to the second order in redshift, yielding the following result:
\[\mathrm{\widetilde{w}_{m}(z)\thickapprox \frac{w_{m,0}}{\alpha^{2\xi}}z^{1-2q_0+2q_1\ln{\alpha}}}\,.\]
In order for models to be feasible, $\mathrm{\widetilde{w}_{m}(z)}$ must be less than or equal to 1, otherwise the symmetric teleparallel density parameter would need to decrease to values that are negative. This sets a lower limit for $\mathrm{q_{1}}$ as follows:
\begin{equation}
{\mathit{ q_{1}\leq \lim_{z\to \infty}-\frac{1}{2} \frac{\left( \ln{\frac{1}{w_{m,0}z}} +2q_0\ln{z}\right)  \left( \alpha-1 \right)}{\left( \alpha-\ln{\frac{1}{z}} +\alpha\ln{\frac{1}{z}} \right)\ln{\alpha} }= \frac{2q_{0}-1}{2\ln{\alpha}}}}\,.
\end{equation}
\begin{table*}[t!]
\caption{\label{Table2}%
The primary outcomes of second application, based on   ($\mathrm{q_0,q_1}$)  presented in \cite{Mamon:2016dlv} }
\begin{ruledtabular}
\begin{tabular*}{\textwidth}{lccccccc}
\textbf{Dataset}                        &$\mathrm{\alpha}$& $\mathrm{q_0}$ & $\mathrm{q_1}$ & $\mathrm{f(Q)/\Lambda}$CDM           & $\mathrm{w_m(z)}\leq 1$          & \textbf{Non-Metricity } & \textbf{Viability} \\
                                        & &       &       & \textbf{compatibility} & \textbf{constraint} & \textbf{EoS}, $\mathrm{w_Q}$           &  \\
\colrule
JLA SNIa+BAO/CMB                       & 2 &  $-0.45$  &  $-2.56$  & not                 & violated                    & diverges\footnote{\label{footnote:2a}When  the matter density parameter exceeds line twice, the symmetric teleparallel EoS undergoes two divergences, as seen in Figs. In particular, \ref{Fig:Mod2}\subref{fig:Mod2-Om} and \ref{Fig:Mod2}\subref{fig:Mod2-wT}.}     & not        \\
JLA SNIa+BAO/CMB                       & 3 &  $-0.54$  &  $-1.35$  & not                 & violated                    & diverges$^\textrm{\ref{footnote:2a}}$     & not        \\
JLA SNIa+BAO/CMB                       & 4 &  $-0.56$  &  $-1.03$  & not                 & violated                    & diverges$^\textrm{\ref{footnote:2a}}$     & not        \\
JLA SNIa+BAO/CMB                       & 5 &  $-0.56$  &  $-0.85$  & not                 & violated                    & diverges$^\textrm{\ref{footnote:2a}}$     & not        \\
\colrule
\multicolumn{8}{l}{\textbf{Utilizing limitation (\ref{q2-min})}}\\
\colrule
JLA SNIa+BAO/CMB                       & 2 &  $-0.45$  &  $-3.366$  & not                 & fulfilled                   & does not diverge & not\footnote{\label{footnote:2b} While the density parameter  matter  constraint (\ref{q2-min}) results in smooth $\mathrm{w_Q}$ patterns Fig. \ref{Fig:Mod2A}\subref{fig:Mod2A-wT}, it fails to generate $\mathrm{f(Q)}$  that is more compatible with $\Lambda$CDM, as indicated in
 Fig. \ref{Fig:Mod2A}\subref{fig:Mod2A-fz}. }       \\
JLA SNIa+BAO/CMB                       & 3 &  $-0.54$  &  $-1.525$  & not                 & fulfilled                   & does not diverge & not$^\textrm{\ref{footnote:2b}}$       \\
JLA SNIa+BAO/CMB                       & 4 &  $-0.56$  &  $-1.114$  & not                 & fulfilled                   & does not diverge & not$^\textrm{\ref{footnote:2b}}$       \\
JLA SNIa+BAO/CMB                       & 5 &  $-0.56$  &  $-0.913$  & not                 & fulfilled                   & does not diverge & not$^\textrm{\ref{footnote:2b}}$
\end{tabular*}
\end{ruledtabular}
\end{table*}
The constrictions ensure that the parametrization (\ref{Mod2-Xz}) is able to approach ${\mathrm q\to 1/2}$ as $\mathrm {z\to \infty}$, as discussed in \cite{Mamon:2016dlv}.  Yet, in reality, this won't stop the density parameter from crossing the unit boundary throughout a wide variety of redshifts; it only smoothes out the peaks at lower redshifts.
  In this case, it may be more important to regulate the amplitude of the $\mathrm{w_{m}(z)}$ peak so that it does not exceed the unit boundary.
   By utilizing Eq.~(\ref{Mod2-Omega_m}), we can determine $\mathrm{q_1}$ by solving this constraint.

\begin{equation}\label{q2-min}
  {\mathit{  q_1=\frac{(\alpha-1)(1+z_{lower})\left[\ln{(1+z_{lower})^{2q_0-1}}-\ln{w_{m,0}}\right]}{\ln\left[\frac{(1+z_l)^{2\left[(\alpha-1)\ln{\alpha}-1\right]
  (1+z_l)}\left(\alpha+z_{lower}\right)^{2(\alpha+z_{lower})}}{\alpha^{2\alpha(1+z_{lower})}}\right]}}}\,.
\end{equation}
It is justifiable to maintain $q_0$ as determined in \cite{Mamon:2016dlv} when the $\mathrm{a_m(z_{lower})}$-peak happens at $\mathrm{z_{lower}}=3$,   for $\alpha$ values between $2 \to 5$, $\mathrm{q_1}$ roughly corresponds to -3.366, -1.525, -1.114, and -0.913, respectively. We chart the progression of the function $\mathrm{f(Q)}$, specifically given by the expression detailed in (\ref{Mod2-f(z)}), as illustrated in Fig. \ref{Fig:Mod2A}\subref{fig:Mod2A-fz}. At high redshifts, we notice discrepancies persist in the plots of $\mathrm{f(Q)}$ relative to anticipated outcomes of $\Lambda$CDM model when compared to Fig. \ref{Fig:Mod2}\subref{fig:Mod2-fz}. Despite the matter density parameter consistently staying below unity, its trend exhibits notable characteristics which does not align with the expected characteristics of a matter-dominated era. This inconsistency is clearly illustrated in Fig. \ref{Fig:Mod2A}\subref{fig:Mod2A-Om}.  Conversely, the universe is unable to accurately depict the  standard CDM matter domination era because $\mathrm{omega^{Tot}> 0}$ is observed at high redshifts, as shown in Fig. \ref{Fig:Mod2A}\subref{fig:Mod2A-weff}. It is important to mention that the improved symmetric teleparallel  equation of state no longer shows any signs of phase transitions, while $\mathrm{\omega_Q}$ now remains finite at all levels of red-shift, as depicted in Fig. \ref{Fig:Mod2A}\subref{fig:Mod2A-wT}. Table \ref{Table2} presents a summary of the outcomes of  second application based on the specified values of $\mathrm{q_0}$ and $\mathrm{q_1}$ presented in \cite{Mamon:2016dlv}, incorporating the matter density restriction (\ref{q2-min}).

\subsection{Third application}\label{Sec4.4}
\begin{figure*}[t!]
\centering
\subfigure[~$\omega^{Tot}$ according to \cite{Mukherjee:2016eqj}]
{\label{fig:Mod3-weff}\includegraphics[scale=.22]{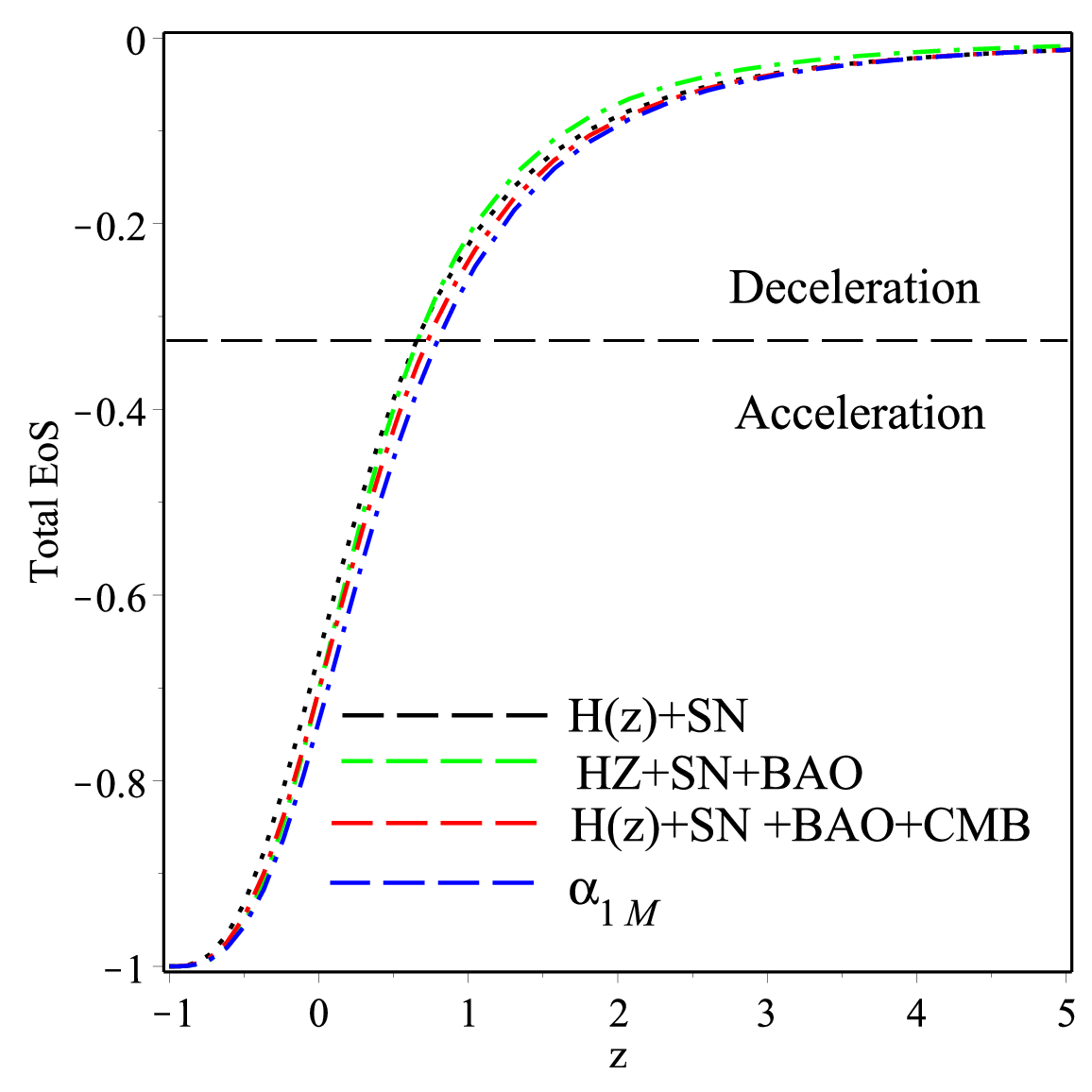}}
\subfigure[~$\omega^{Tot}$ according to \cite{Mukherjee:2016eqj}]
{\label{fig:Mod3-fz}\includegraphics[scale=.22]{{{JFRMMM_3FAW_Mod3_fz}}}}
\subfigure[~$w_{m}(z)$, $w_{Q}(z)$ according to \cite{Mukherjee:2016eqj}]
{\label{fig:Mod3-Om}\includegraphics[scale=.22]{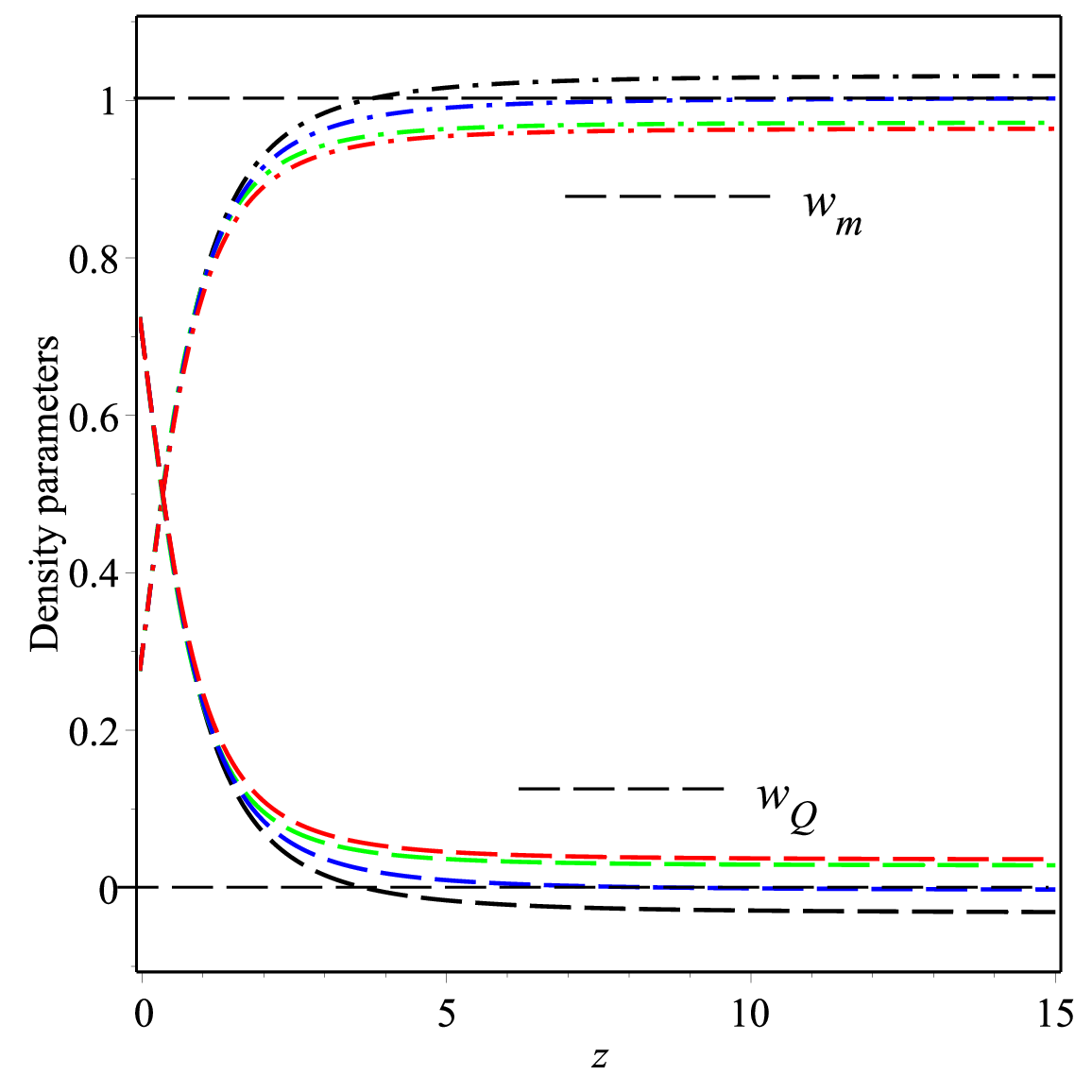}}
\subfigure[~$\omega_{Q}$ according to \cite{Mukherjee:2016eqj}]
{\label{fig:Mod3-wT}\includegraphics[scale=.22]{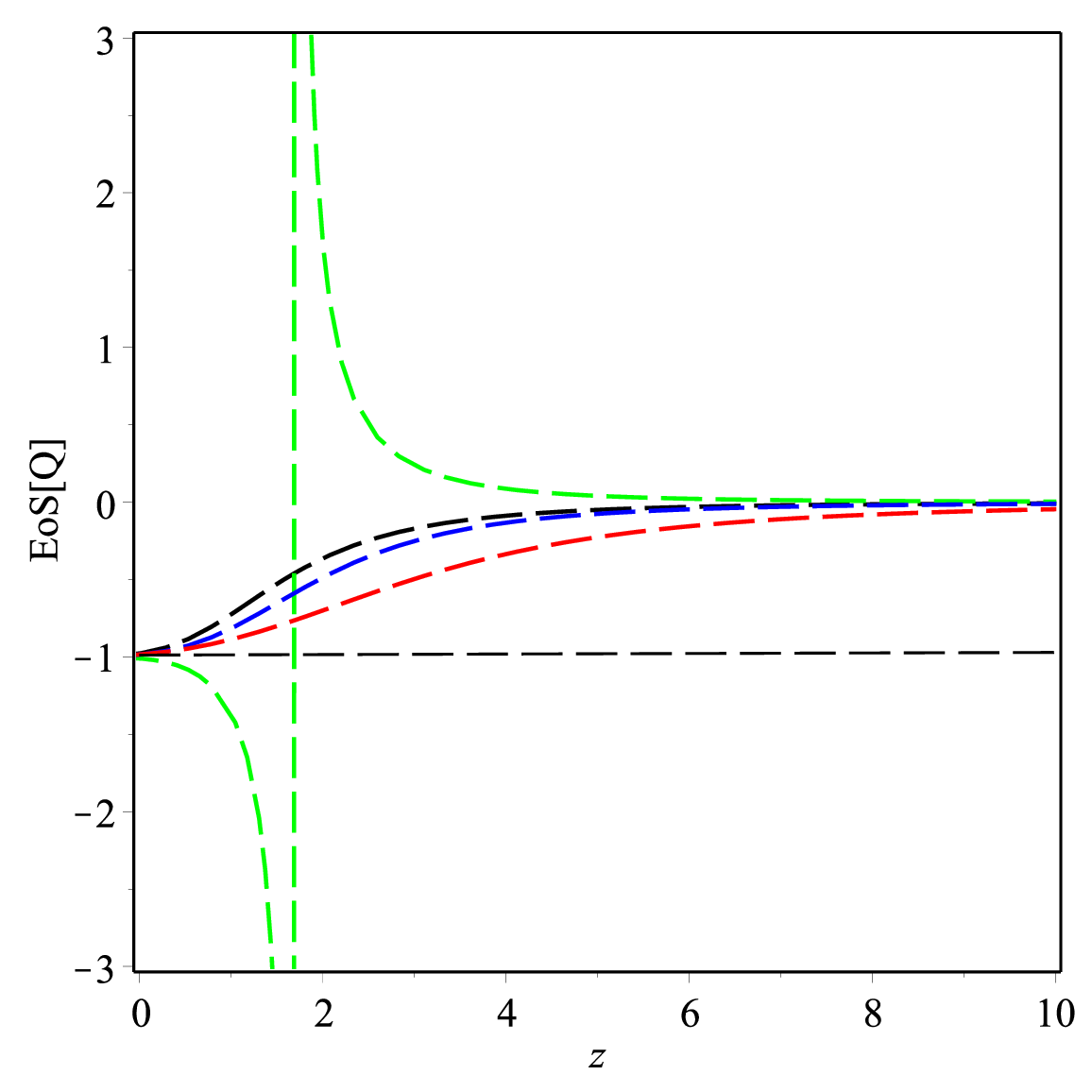}}
\subfigure[~$\omega^{Tot}$ using Eq.~(\ref{alpha-min})]
{\label{fig:Mod3A-weff1}\includegraphics[scale=.22]{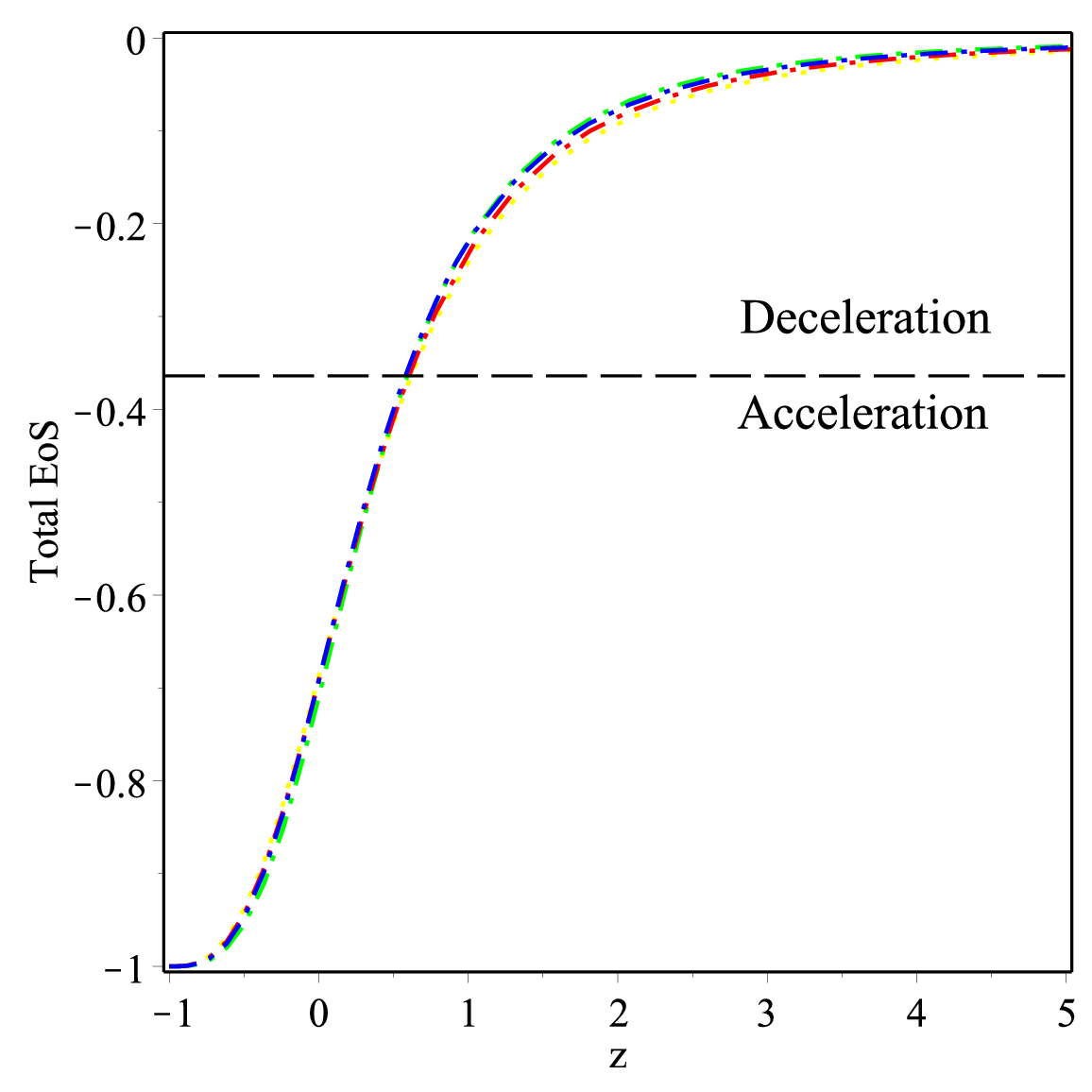}}
\subfigure[~$w_{m}$, $w_{Q}$ using  Eq.~(\ref{alpha-min})]
{\label{fig:Mod3A-Om1}\includegraphics[scale=.22]{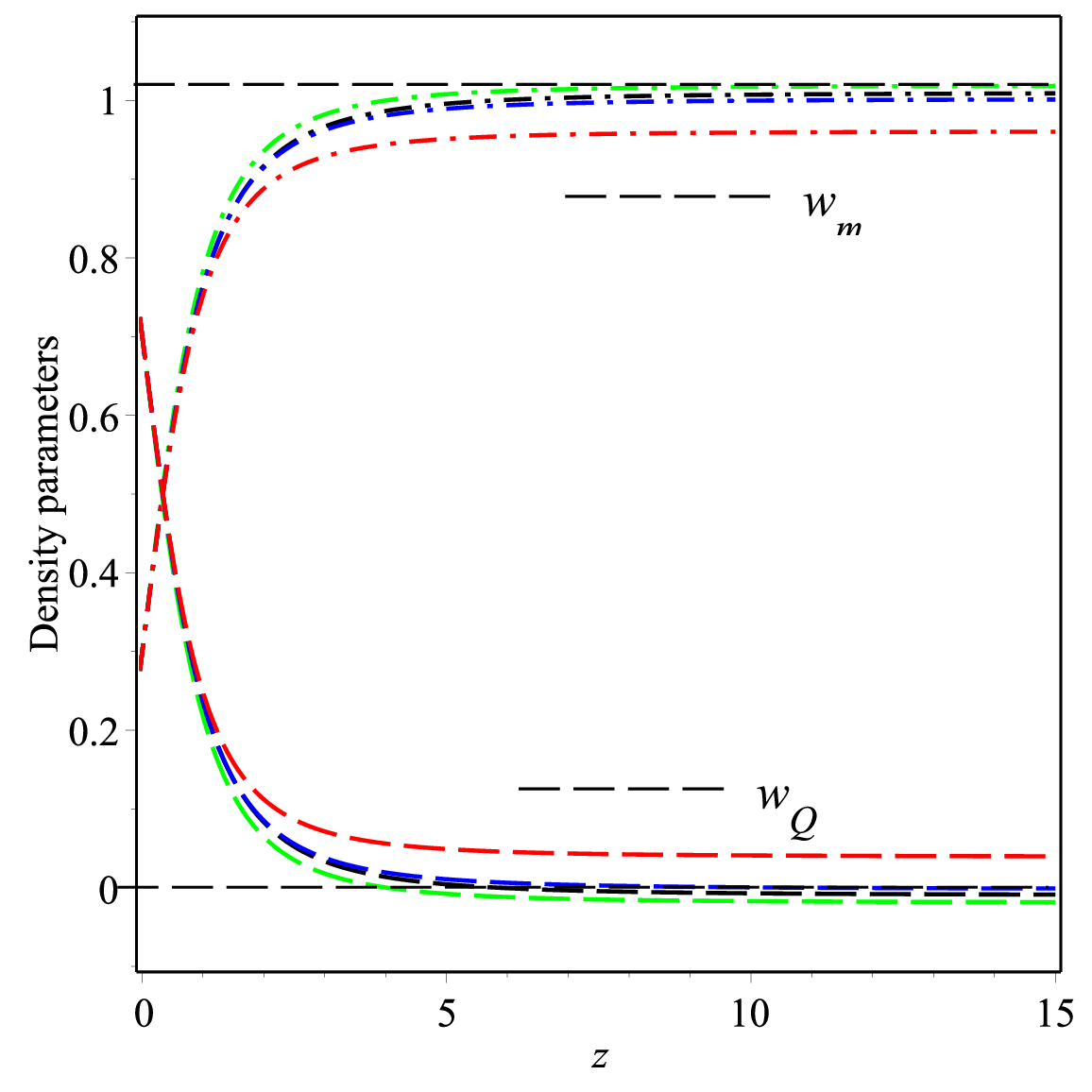}}
\subfigure[~$w_{Q}$ using  Eq.~(\ref{alpha-min})]
{\label{fig:Mod3A-wT1}\includegraphics[scale=.22]{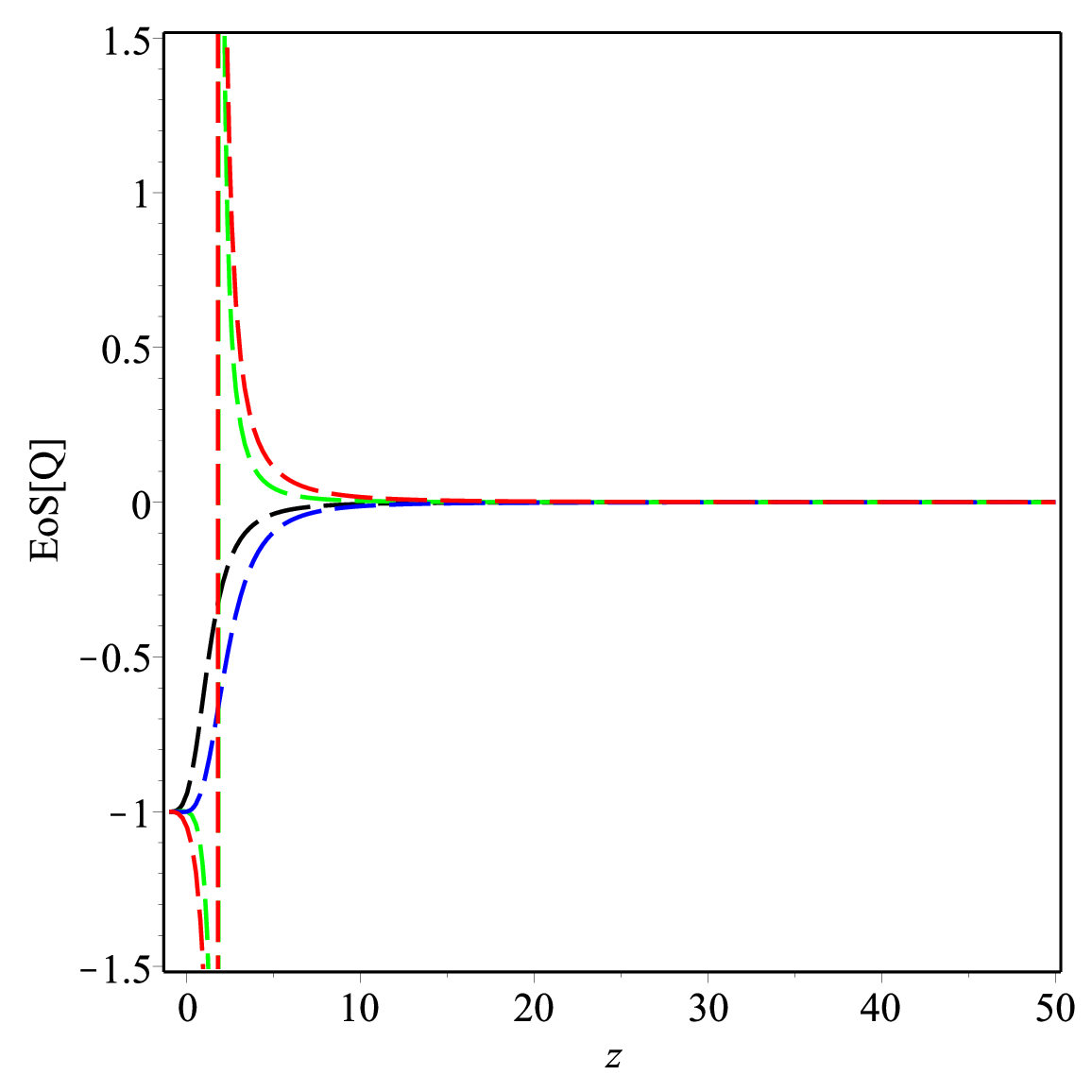}}
\caption[figtopcap]{The optimal values for the model parameters  ($\alpha_1$, $\alpha_2$) are determined based on the dataset combination \cite{Mukherjee:2016eqj} as assessed in subfigures \subref{fig:Mod3-weff} -- \subref{fig:Mod3-wT}.
 In sub figures \subref{fig:Mod3-weff} -- \subref{fig:Mod3-wT},  the additive constant is assumed to be $\delta \alpha_1=10^{-3}$, we apply the constraint (\ref{alpha-min}) to find $\alpha_{min}$ for various options of $\alpha_2$.. {   The labels used in Fig.~\ref{Fig:Mod3}\subref{fig:Mod3-weff} are the same in Figs.~\ref{Fig:Mod3}\subref{fig:Mod3-fz}--\ref{Fig:Mod3}\subref{fig:Mod3A-wT1}.}}
\label{Fig:Mod3}
\end{figure*}

Now we are going to rebuild the theory of $\mathrm{f(Q)}$ gravity based on the parametric representation of the effective equation of state previously used in Ref.~\cite{Mukherjee:2016eqj},
\begin{equation}\label{Mod3-weff}
    {\omega^{Tot}=-\frac{1}{1+\alpha_1(1+z)^{\alpha_2}}}\,,
\end{equation}
where the two parameters of the model are denoted by $\alpha_1$ and $\alpha_2$. At high redshifts, the universe essentially creates standard CDM as $\omega^{Tot}$ approaches zero. As $z$ approaches -1, the universe essentially moves towards de Sitter when $\omega^{Tot}$ approaches -1. This is clearly illustrated in Fig.~\ref{Fig:Mod3}\subref{fig:Mod3-weff}. Essentially, this design has the potential to create a thriving cosmic timeline. By utilizing (\ref{deceleration}), we can express the deceleration parameter related to the parametrization mentioned above as:
\begin{equation}\label{Mod3-qz}
\mathrm{   q(z)=-1+\frac{3\alpha_1(1+z)^{\alpha_2}}{2\left[1+\alpha_1(1+z)^{\alpha_2}\right]}}\,.
\end{equation}
Plugging the above data into equation (\ref{Hubble-deceleration}), then $H(z)$ is determined as:
\begin{equation}\label{Mod3-H}
 {\mathit{   H(z)=H_{0}\left[\frac{1+\alpha_1(1+z)^{\alpha_2}}{1+\alpha_1}\right]^{\frac{3}{2\alpha_2}}}}\,.
\end{equation}
It is evident that when $\alpha_2=3$, the model results in the $\Lambda$CDM model. In this case, we find that $w_{m,0}=\frac{\alpha_1}{1+\alpha_1}$, which is also expressed as $\alpha_1=\frac{w_{m,0}}{1-w_{m,0}}$.  The value $w_{m,0}=0.297$ is adopted from the measurement in Ref. \cite{Mukherjee:2016eqj}. In dynamic DE models that are feasible, it is likely that the value of $\alpha_2$ will be around $3$. Alternatively, we employ equation (\ref{Reconstruction2}) to calculate $\mathrm{f(Q)}$ gravity, leading to the parametric expression (\ref{Mod3-weff}) we get:
\begin{eqnarray}
\nonumber    f(z)&=&-9\alpha_1\Omega_{m,0}H_0^2\left[1+\alpha_1(1+z)^{\alpha_2}\right]^{\frac{3}{2\alpha_2}}    \int_0^z{\frac{(1+\tilde{z})^{\alpha_2+2}}{\left[1+\alpha_1(1+\tilde{z})^{\alpha_2}\right]^{1+\frac{3}{2\alpha_2}}}}d\tilde{z}.\label{Mod3-f(z)}
\end{eqnarray}
A comparative assessment of the $\mathrm{f(Q)}$ gravity model against the $\Lambda$CDM framework for varying parameter settings within the model $\alpha_1$ and $\alpha_2$, based on the dataset from \cite{Mukherjee:2016eqj}, is illustrated in Fig. \ref{Fig:Mod3} \subref{fig:Mod3-fz}. Utilizing the combined dataset of SN+observed Hubble data (OHD), the plots reveal a consistent deviation of the $\mathrm{f(Q)}$ gravity model from $\Lambda$CDM, as it does not exhibit oscillations around the $\Lambda$CDM curve. Conversely, $\Lambda$CDM remains consistent with the theory in other scenarios. The reasons behind these findings will be explored in subsequent sections of this study.

Employing the deceleration parameter from equation (\ref{Mod3-qz}), the matter density parameter in equation (\ref{matter-density-parameter}) is expressed as:
\begin{equation}\label{Mod3-Omega_m}
  {\mathit { w_{m}(z)=w_{m,0}\,(1+z)^{3}\left[\frac{1+\alpha_1}{1+\alpha_1(1+z)^{\alpha_2}}\right]^{\frac{3}{\alpha_2}}}}.
\end{equation}
We explore a range of model parameter values to illustrate the evolution of the matter density over tim in Fig. \ref{Fig:Mod3}\subref{fig:Mod3-Om}. It is demonstrable that by appropriately tuning the parameters, the model can be made to accurately fit the observed data of SN+OHD dataset, the matter density parameter surpasses 1 at a redshift of approximately $z \sim 2$. However, upon aligning the parameters to match the dataset, it becomes evident how closely the model can replicate the observed phenomena $H(z)+SN$, the density matter shows a small but significant departure from $\Lambda$CDM at high $z$.  Nevertheless, when the BAO and CMB (shift parameters) are included, it evolves somewhat similarly to $\Lambda$CDM. The plots of Fig. \ref{Fig:Mod3} \subref{fig:Mod3-fz} actually match these results.

We assess the symmetric teleparallel (DE) EoS by substituting from (\ref{Mod3-H}) and (\ref{Mod3-f(z)}) in (\ref{wQ(z)}) as follows:
\begin{equation}\label{Mod3-wDE}
    {\mathit { w_{Q}=-\frac{\left[1+\alpha_1(1+z)^{\alpha_2}\right]^{\frac{3}{{\alpha_2}}}-\alpha_1(1+z)^{\alpha_2} \left[1+\alpha_1(1+z)^{\alpha_2}\right]^{\frac{3}{{\alpha_2}}-1}}{\left[1+\alpha_1(1+z)^{\alpha_2}\right]^{\frac{3}{{\alpha_2}}}-w_{m,0}(1+z)^3 (1+\alpha_1)^{\frac{3}{{\alpha_2}}}}}}.
\end{equation}
Fig. \ref{Fig:Mod3}\subref{fig:Mod3-wT} shows that, when using the dataset SN+OHD+BAO, the symmetric teleparallel EoS diverges at red-shift $z\sim 2$. In Figure \ref{Fig:Mod3}\subref{fig:Mod3-Om}, it is illustrated that the matter density parameter breaching the unit threshold is coincident with the phase transition experienced within the framework of symmetric teleparallel gravity.
\begin{table*}[t!]
\caption{\label{Table3}%
Utilizing the constraint on the matter density parameter given by equation (\ref{alpha-min}) and the values of ($\alpha_1,~\alpha_2$) parameters as stated in Ref. \cite{Mukherjee:2016eqj}, the primary outcomes of model 3 are as follows: $\alpha_1=\alpha_{1\, min}+\delta \alpha_1(=10^{-3})$. As determined in \cite{Mukherjee:2016eqj}, we use $\Omega_{m,0}=0.297$ in all treatments.}
\begin{ruledtabular}
\begin{tabular}{lcccccc}
\textbf{Dataset}            & $\alpha_1$ & $\alpha_2$ & $f(Q)/\Lambda$CDM           & $w_m(z)\leq 1$          & \textbf{symmetric teleparallel} & \textbf{Viability} \\
                            &          &       & \textbf{compatibility} & \textbf{constraint} & \textbf{EoS}, $w_Q$           &  \\
\colrule
SN+OHD                      &   $0.445$\footnote{\label{footnote:3a}It is important to note that this dataset yields $\alpha_1 < \alpha_{1\, \text{min}}$, which accounts for the breach of the matter density parameter constraint.}   & $2.8$  & not                & violated                  & diverges     & not        \\
SN+OHD+BAO                  &   $0.409$    & $3.13$  & semi               & fulfilled                 & does not diverge     & yes        \\
SN+OHD+BAO+CMB              &   $0.444$    & $2.907$  & semi             & fulfilled                 & does not diverge     & yes        \\
\colrule
\multicolumn{7}{l}{\textbf{Applying Eq. (\ref{alpha-min})}}\\
\colrule
case (i)                      &   $0.503$    & $2.7 \lesssim 3$  & semi               & fulfilled                   & doesnt diverge   & Ok.      \\
case (ii)                  &   $0.356$    & $3.3 \gtrsim3$  & semi               & fulfilled                   & does not diverge (quintom) & yes      \\
case (iii)              &   $0.421$    & $3$  & semi               & fulfilled                   & doesnt diverge   & Ok.      \\
\colrule
$\Lambda$CDM                            & $\alpha_{1\,min}$   & 3 & yes                & fulfilled                   & $-1$           & yes
\end{tabular}
\end{ruledtabular}
\end{table*}

As outlined in Section \ref{Sec4.1}, we impose constraints on the model parameters to guarantee that the matter density parameter (\ref{Mod3-Omega_m}) gradually nears its maximum value $\Omega_{m,0}=1$ in the limit where $z$ tends to infinity. This constraint assists in setting a minimum threshold for the parameter $\alpha$. The leading term in the asymptotic expression for the matter density parameter is provided by: $$\widetilde{w}_{m}(z)\thickapprox w_{m,0} \left(1+\frac{1}{\alpha_1}\right)^{\frac{3}{\alpha_2}}.$$
$\widetilde{w}_{m}(z)\leq 1$ is required for viable models; otherwise, the symmetric teleparallel the density parameter would drop below zero, establishing a lower threshold for $\alpha_1$
\begin{equation}\label{alpha-min}
 {\mathit{  \alpha_{1\,min}=\frac{w_{m,0}^{\frac{\alpha_2}{3}}}{1-w_{m,0}^{\frac{\alpha_2}{3}}}}}.
\end{equation}
Should $\alpha_1$ fall short of the designated minimal value, the matter density parameter would surpass 1 at a certain redshift in the past, thereby leading to the torsion density parameter becoming negative. For instance, analysis of the SN+OHD dataset \cite{Mukherjee:2016eqj} indicates that the estimated $\alpha_1$ value of $0.445$ is below the necessary minimum $\alpha_{1, \text{min}} = 0.4728$, which is required to keep the matter density within acceptable limits. Importantly, we utilize $w_{m,0}=0.297$ as reported in \cite{Mukherjee:2016eqj} and $\alpha_2=2.8$ as derived from the SN+OHD dataset (\ref{alpha-min}). Applying the SN+OHD dataset reveals a conflict between model predictions and observed data, as depicted in Figure \ref{Fig:Mod3}\subref{fig:Mod3-Om}. Conversely, when incorporating the SN+OHD+BAO dataset, we discover that the observed values of $\alpha_1=0.409 > \alpha_{1,min}=0.3904$ and $\alpha_1=0.444 > \alpha_{1,min}=0.4438$ align and allow for coherent cosmological models.

Notably, for the $\Lambda$CDM scenario ($\alpha_2=3$), the minimal value of $\alpha_1$ corresponds directly to the proportion of the matter density parameter to the symmetric teleparallel density parameter at the current epoch. Specifically, $\alpha_{1, \text{min}} = \frac{w_{m,0}}{w_{\lambda,0}}$. In the illustrations provided in Figures \ref{Fig:Mod3}\subref{fig:Mod3A-weff1}--\subref{fig:Mod3A-wT1}, we present the $\Lambda$CDM model with $n=3$ and $\alpha_1$ adjusted to its minimum allowable value, $\alpha_{1, \text{min}}$. These configurations highlight the interplay between the matter and teleparallel components within the model to find a fixed symmetric teleparallel EoS $w_Q=-1$. We describe three potential scenarios that could work if $\alpha_1$ marginally surpasses $\alpha_{1\,min}$:
(i) We find that $w_Q$ behaves in a quintessence-like manner, with $w_Q$ currently being slightly greater than $-1$ for values of $\alpha_2$ less than or approximately equal to 3.
(ii) The equation of state parameter $w_Q$ transitions from quintessence to phantom behavior, crossing the phantom dividing line at a redshift of approximately 4, with $w_Q$ currently being less than $-1$. This suggests a quintom-like behavior for the symmetric teleparallel equation of state when $n$ is greater than or approximately equal to 3.
When $\alpha_2$ is set to 3, the symmetric teleparallel  EoS  progresses through a quintessence phase with $w_Q$ approximately equal to $-1$. At very high redshifts, $w_Q$ consistently approaches 0, which accounts for the late-time accelerated expansion. In the future, it evolves towards a state resembling a cosmological constant, representing a pure de Sitter universe. The outcomes of the model are summarized in Table \ref{Table3}.

\section{Concluding remarks}\label{sec6}

Recently, parametric representations of the deceleration parameter have been investigated as a means to kinematically explain the universe's accelerated expansion at late times. Although numerous parametric models can effectively describe the shift from a decelerating to an accelerating phase, it is crucial to analyze these models within dynamic frameworks or theories of modified gravity. Such an approach enables the evaluation of cosmological parameters not only at the foundational level but also in terms of perturbations, providing a comprehensive test of their validity.

In this study, we develop a reconstruction method for $f\mathrm{(Q)}$ gravity that accommodates any specific $\mathrm{q(z)}$. This is achieved by ensuring consistency between the deceleration parameter (\ref{Hubble-deceleration}) and the $\mathrm{f(Q)}$ gravity framework (\ref{phase-portrait-z}). Additionally, using information about either   $\mathrm{\omega_{DE}(z)}$  or   $\mathrm{\omega^{Tot}(z)}$, we derive two further reconstruction equations.

We tested three distinct applications in this study.\\

 First Application: We used the $\mathrm{q(z)}$ parametrization from Eq.~(\ref{Mod1-Xz}) and compared the $\mathrm{f(Q)}$  gravity model with $\Lambda$CDM   The results show that the model is not viable, even with additional parameters. \\

Second Application: We employed the $\mathrm{q(z)}$ parametrization from Eq.~(\ref{Mod2-Xz}). Similar to the first application, this model fails to produce a viable cosmic scenario consistent with $\Lambda$CDM.\\
Third Application: We used the $\omega^{Tot}(z)$ parametrization from Eq.~(\ref{Mod3-weff}). Here, the $\mathrm{f(Q)}$  gravity model aligns well with current observational data and $\Lambda$CDM predictions. This model is expected to provide insights into dark energy properties beyond  $\Lambda$CDM as it can generate dynamic dark energy behavior, such as quintessence or quintom.

In each of the three scenarios, the teleparallel EoS exhibits divergence whenever the matter density parameter surpasses 1 or if the symmetric teleparallel density parameter turns negative. This characteristic offers a potential avenue for verification by future dark energy surveys.
In summary, $\mathrm{f(Q)}$ modified gravity provides a promising framework for understanding cosmic acceleration and addressing long-standing cosmological challenges through reconstruction techniques. However, a more comprehensive study would also explore the exploration of the theory at the level of perturbations. We reserve this task for future investigation.

%

%
%

\end{document}